\documentclass[longauth]{aa}      
\usepackage{placeins}
\usepackage{amsmath}
\usepackage{graphicx}
\usepackage{txfonts}
\usepackage{hyperref}

\hypersetup{
    colorlinks=true,
    citecolor=blue,
    linkcolor=blue,
    urlcolor=blue,
}

\makeatletter
\renewcommand*\aa@pageof{, page \thepage{} of \pageref*{LastPage}}
\makeatother

\usepackage{color}
\usepackage{xcolor}
\newcommand{\valerio}[1]{#1}

\begin{document} 

   \title{A new dynamical modeling of the WASP-47 \\ system with CHEOPS observations\thanks{This article uses data from CHEOPS programs \texttt{CH\_PR100017} and \texttt{CH\_PR100025}. The individual data sets are listed in Table~\ref{tab:datasets}.}}
   \authorrunning{V. Nascimbeni et al.}
   \titlerunning{A new dynamical modeling of WASP-47}


   \author{
V. Nascimbeni\inst{1}\thanks{send offprint requests to: \texttt{valerio.nascimbeni@inaf.it}}, 
L. Borsato\inst{1}, 
T. Zingales\inst{2,1}, 
G. Piotto\inst{2,1}, 
I. Pagano\inst{3}, 
M. Beck\inst{4}, 
C. Broeg\inst{5,6}, 
D. Ehrenreich\inst{4,7}, 
S. Hoyer\inst{8}, 
F. Z. Majidi\inst{2,1}, 
V. Granata\inst{9,1}, 
S. G. Sousa\inst{10}, 
T. G. Wilson\inst{11}, 
V. Van Grootel\inst{12}, 
A. Bonfanti\inst{13}, 
S. Salmon\inst{4}, 
A. J. Mustill\inst{14,52}, 
L. Delrez\inst{15,12}, 
Y. Alibert\inst{5}, 
R. Alonso\inst{16,17}, 
G. Anglada\inst{18,19}, 
T. Bárczy\inst{20}, 
D. Barrado\inst{21}, 
S. C. C. Barros\inst{10,22}, 
W. Baumjohann\inst{13}, 
T. Beck\inst{5}, 
W. Benz\inst{5,6}, 
M. Bergomi\inst{1}, 
N. Billot\inst{4}, 
X. Bonfils\inst{23}, 
A. Brandeker\inst{24}, 
J. Cabrera\inst{25}, 
S. Charnoz\inst{26}, 
A. Collier Cameron\inst{11}, 
Sz. Csizmadia\inst{25}, 
P. E. Cubillos\inst{27,13}, 
M. B. Davies\inst{28}, 
M. Deleuil\inst{8}, 
A. Deline\inst{4}, 
O. D. S. Demangeon\inst{10,22}, 
B.-O. Demory\inst{6,5}, 
A. Erikson\inst{25}, 
A. Fortier\inst{5,6}, 
L. Fossati\inst{13}, 
M. Fridlund\inst{29,30}, 
D. Gandolfi\inst{31}, 
M. Gillon\inst{15}, 
M. Güdel\inst{32}, 
K. G. Isaak\inst{33}, 
L. L. Kiss\inst{34,35}, 
J. Laskar\inst{36}, 
A. Lecavelier des Etangs\inst{37}, 
M. Lendl\inst{4}, 
C. Lovis\inst{4}, 
R. Luque\inst{38,39}, 
D. Magrin\inst{1}, 
P. F. L. Maxted\inst{40}, 
C. Mordasini\inst{6}, 
G. Olofsson\inst{24}, 
R. Ottensamer\inst{41}, 
E. Pallé\inst{16}, 
G. Peter\inst{42}, 
D. Piazza\inst{5}, 
D. Pollacco\inst{43}, 
D. Queloz\inst{44,45}, 
R. Ragazzoni\inst{1,2}, 
N. Rando\inst{46}, 
F. Ratti\inst{46}, 
H. Rauer\inst{25,47,48}, 
I. Ribas\inst{18,19}, 
N. C. Santos\inst{10,22}, 
G. Scandariato\inst{3}, 
D. Ségransan\inst{4}, 
A. E. Simon\inst{5}, 
A. M. S. Smith\inst{25}, 
M. Steinberger\inst{13}, 
M. Steller\inst{13}, 
Gy. M. Szabó\inst{49,50}, 
N. Thomas\inst{5}, 
S. Udry\inst{4}, 
J. Venturini\inst{4}, 
N. A. Walton\inst{51}, 
D. Wolter\inst{25}
          }

   \institute{
\label{inst:1} INAF, Osservatorio Astronomico di Padova, Vicolo dell'Osservatorio 5, 35122 Padova, Italy \and
\label{inst:2} Dipartimento di Fisica e Astronomia "Galileo Galilei", Universita degli Studi di Padova, Vicolo dell'Osservatorio 3, 35122 Padova, Italy \and
\label{inst:3} INAF, Osservatorio Astrofisico di Catania, Via S. Sofia 78, 95123 Catania, Italy \and
\label{inst:4} Observatoire Astronomique de l'Université de Genève, Chemin Pegasi 51, CH-1290 Versoix, Switzerland \and
\label{inst:5} Physikalisches Institut, University of Bern, Sidlerstrasse 5, 3012 Bern, Switzerland \and
\label{inst:6} Center for Space and Habitability, University of Bern, Gesellschaftsstrasse 6, 3012 Bern, Switzerland \and
\label{inst:7} Centre Vie dans l’Univers, Facult\'e des sciences, Universit\'e de Gen\'eve, Quai Ernest-Ansermet 30, CH-1211 Gen\'eve 4, Switzerland \and
\label{inst:8} Aix Marseille Univ, CNRS, CNES, LAM, 38 rue Frédéric Joliot-Curie, 13388 Marseille, France \and
\label{inst:9} Centro di Ateneo di Studi e Attività Spaziali "Giuseppe Colombo" (CISAS), Università degli Studi di Padova, Via Venezia 15, 35131 Padova, Italy \and
\label{inst:10} Instituto de Astrofisica e Ciencias do Espaco, Universidade do Porto, CAUP, Rua das Estrelas, 4150-762 Porto, Portugal \and
\label{inst:11} Centre for Exoplanet Science, SUPA School of Physics and Astronomy, University of St Andrews, North Haugh, St Andrews KY16 9SS, UK \and
\label{inst:12} Space sciences, Technologies and Astrophysics Research (STAR) Institute, Université de Liège, Allée du 6 Août 19C, 4000 Liège, Belgium \and
\label{inst:13} Space Research Institute, Austrian Academy of Sciences, Schmiedlstrasse 6, A-8042 Graz, Austria \and
\label{inst:14} Division of Astrophysics, Department of Physics, Lund University, Box 43, 221 00 Lund, Sweden \and
\label{inst:15} Astrobiology Research Unit, Université de Liège, Allée du 6 Août 19C, B-4000 Liège, Belgium \and
\label{inst:16} Instituto de Astrofisica de Canarias, 38200 La Laguna, Tenerife, Spain \and
\label{inst:17} Departamento de Astrofisica, Universidad de La Laguna, 38206 La Laguna, Tenerife, Spain \and
\label{inst:18} Institut de Ciencies de l'Espai (ICE, CSIC), Campus UAB, Can Magrans s/n, 08193 Bellaterra, Spain \and
\label{inst:19} Institut d'Estudis Espacials de Catalunya (IEEC), 08034 Barcelona, Spain \and
\label{inst:20} Admatis, 5. Kandó Kálmán Street, 3534 Miskolc, Hungary \and
\label{inst:21} Depto. de Astrofisica, Centro de Astrobiologia (CSIC-INTA), ESAC campus, 28692 Villanueva de la Cañada (Madrid), Spain \and
\label{inst:22} Departamento de Fisica e Astronomia, Faculdade de Ciencias, Universidade do Porto, Rua do Campo Alegre, 4169-007 Porto, Portugal \and
\label{inst:23} Université Grenoble Alpes, CNRS, IPAG, 38000 Grenoble, France \and
\label{inst:24} Department of Astronomy, Stockholm University, AlbaNova University Center, 10691 Stockholm, Sweden \and
\label{inst:25} Institute of Planetary Research, German Aerospace Center (DLR), Rutherfordstrasse 2, 12489 Berlin, Germany \and
\label{inst:26} Université de Paris, Institut de physique du globe de Paris, CNRS, F-75005 Paris, France \and
\label{inst:27} INAF, Osservatorio Astrofisico di Torino, Via Osservatorio, 20, I-10025 Pino Torinese To, Italy \and
\label{inst:28} Centre for Mathematical Sciences, Lund University, Box 118, 221 00 Lund, Sweden \and
\label{inst:29} Leiden Observatory, University of Leiden, PO Box 9513, 2300 RA Leiden, The Netherlands \and
\label{inst:30} Department of Space, Earth and Environment, Chalmers University of Technology, Onsala Space Observatory, 439 92 Onsala, Sweden \and
\label{inst:31} Dipartimento di Fisica, Universita degli Studi di Torino, via Pietro Giuria 1, I-10125, Torino, Italy \and
\label{inst:32} University of Vienna, Department of Astrophysics, Türkenschanzstrasse 17, 1180 Vienna, Austria \and
\label{inst:33} Science and Operations Department - Science Division (SCI-SC), Directorate of Science, European Space Agency (ESA), European Space Research and Technology Centre (ESTEC),
Keplerlaan 1, 2201-AZ Noordwijk, The Netherlands \and
\label{inst:34} Konkoly Observatory, Research Centre for Astronomy and Earth Sciences, 1121 Budapest, Konkoly Thege Miklós út 15-17, Hungary \and
\label{inst:35} ELTE E\"otv\"os Lor\'and University, Institute of Physics, P\'azm\'any P\'eter s\'et\'any 1/A, 1117 Budapest, Hungary \and
\label{inst:36} IMCCE, UMR8028 CNRS, Observatoire de Paris, PSL Univ., Sorbonne Univ., 77 av. Denfert-Rochereau, 75014 Paris, France \and
\label{inst:37} Institut d'astrophysique de Paris, UMR7095 CNRS, Université Pierre \& Marie Curie, 98bis blvd. Arago, 75014 Paris, France \and
\label{inst:38} Instituto de Astrofísica de Andalucía, Consejo Superior de Investigaciones Científicas, 18008 Granada, Spain \and
\label{inst:39} Department of Astronomy and Astrophysics, University of Chicago, Chicago, IL 60637, USA \and
\label{inst:40} Astrophysics Group, Keele University, Staffordshire, ST5 5BG, United Kingdom \and
\label{inst:41} Department of Astrophysics, University of Vienna, Tuerkenschanzstrasse 17, 1180 Vienna, Austria \and
\label{inst:42} Institute of Optical Sensor Systems, German Aerospace Center (DLR), Rutherfordstrasse 2, 12489 Berlin, Germany \and
\label{inst:43} Department of Physics, University of Warwick, Gibbet Hill Road, Coventry CV4 7AL, United Kingdom \and
\label{inst:44} ETH Zurich, Department of Physics, Wolfgang-Pauli-Strasse 2, CH-8093 Zurich, Switzerland \and
\label{inst:45} Cavendish Laboratory, JJ Thomson Avenue, Cambridge CB3 0HE, UK \and
\label{inst:46} ESTEC, European Space Agency, 2201AZ, Noordwijk, NL \and
\label{inst:47} Zentrum für Astronomie und Astrophysik, Technische Universität Berlin, Hardenbergstr. 36, D-10623 Berlin, Germany \and
\label{inst:48} Institut für Geologische Wissenschaften, Freie Universität Berlin, 12249 Berlin, Germany \and
\label{inst:49} ELTE E\"otv\"os Lor\'and University, Gothard Astrophysical Observatory, 9700 Szombathely, Szent Imre h. u. 112, Hungary \and
\label{inst:50} MTA-ELTE Exoplanet Research Group, 9700 Szombathely, Szent Imre h. u. 112, Hungary \and
\label{inst:51} Institute of Astronomy, University of Cambridge, Madingley Road, Cambridge, CB3 0HA, United Kingdom \and
\label{inst:52} Lund Observatory, Dept. of Astronomy and Theoretical Physics, Lund University, Box 43, 22100 Lund, Sweden 
}
   \date{Submitted: 2022, November 11 --- Accepted: 2023, February 2}

  \abstract{
  Among the hundreds of known hot Jupiters (HJs), only five have been found to have companions on short-period orbits. Within this rare class of multiple planetary systems, the architecture of WASP-47 is unique, hosting an HJ (planet -b) with both an inner and an outer sub-Neptunian mass companion (-e and -d, respectively) as well as an additional non-transiting, long-period giant (-c). The small period ratio between planets -b and -d boosts the transit time variation (TTV) signal, making it possible to reliably measure the masses of these planets in synergy with the radial velocity (RV) technique. In this paper, we present new space- and ground-based photometric data of WASP-47b and WASP-47-d, including 11 unpublished light curves from the ESA mission CHaracterising ExOPlanet Satellite (CHEOPS). We analyzed the light curves in a homogeneous way together with all the publicly available data to carry out a global $N$-body dynamical modeling of the TTV and RV signals. We retrieved, among other parameters, a mass and density for planet -d  of $M_\mathrm{d}=15.5\pm 0.8$~$M_\oplus$ and $\rho_\mathrm{d}=1.69\pm 0.22$~g\,cm$^{-3}$, which is in good agreement with the literature and consistent with a Neptune-like composition. For the inner planet (-e), we found a mass and density of $M_\mathrm{e}=9.0\pm 0.5$~$M_\oplus$ and $\rho_\mathrm{e}=8.1\pm 0.5$~g\,cm$^{-3}$, suggesting an Earth-like composition close to other ultra-hot planets at similar irradiation levels. Though this result is in agreement with previous RV plus TTV studies, it is not in agreement with the most recent RV analysis (at 2.8$\sigma$), which yielded a lower density compatible with a pure silicate composition.  This discrepancy highlights the still unresolved issue of suspected systematic offsets between RV and TTV measurements. In this paper, we also significantly improve the orbital ephemerides of all transiting planets, which will be crucial for any future follow-up.
  }


   \maketitle
%
\section{Introduction}

Hot Jupiters (HJs), that is, giant, gaseous planets ($0.3\lesssim M_\mathrm{p} \lesssim 13$~$M_\mathrm{jup}$) orbiting their host stars on very short orbits ($P\lesssim 10$~d), have been the subject of many ``firsts'' in the history of exoplanetary research, as the first planet discovered around a solar-type star (51~Peg~b; \citealt{MAYOR&QUELOZ1995}) is an HJ and the first reported  transiting planet (HD~209458~b; \citealt{Charbonneau2000,Henry2000}), which is also the first planet where an atmospheric feature has been detected (\citealt{Charbonneau2002}), is an HJ. Although 479 HJs have been discovered so far by various techniques,\footnote{Source: NASA Exoplanet Archive. \url{https://exoplanetarchive.ipac.caltech.edu/}.} albeit mostly by combining transit photometry and radial velocity (RV) measurements, many key questions are still unanswered about their formation, migration, and successive evolution \citep{Dawson2018,Fortney2021}.

\begin{table*}
    \centering\renewcommand{\arraystretch}{1.2}
    \caption{Mass measurements of the WASP-47 planets published in the literature.}
    \begin{tabular}{l|cccc|c} \hline\hline
       Reference  & $M_\textrm{e}/M_\oplus$ & $M_\textrm{b}/M_\oplus$ & $M_\textrm{d}/M_\oplus$ & $M_\textrm{c}\sin i/M_\oplus$ & Method \\ \hline
        \citet{Hellier2012} & --- & $364\pm 16$ & --- & --- & RV \rule{0pt}{14pt}\\ 
        \citet{NeveuVanmalle2016} & --- & $360\pm 19$ & --- & $396\pm 70$ & RV \\ 
        \citet{Dai2015} & $12.2\pm 3.7$ & $370\pm 29$ & $10.4\pm 8.4^\dagger$ & --- & RV \\ 
        \citet{Becker2015} & $<22^\ddagger$ & $341^{+73}_{-55}$ & $15 \pm 7$ & --- & RV+TTV \\ 
        \citet{Almenara2016} & $9.1^{+1.8}_{-2.9}$ & $364\pm 8$ & $15.7\pm 1.1$ & $361^{+80}_{-54}$ & RV+TTV$^\mathparagraph$ \\ 
        \citet{Weiss2017} & $9.1\pm 1.0$ & $358\pm 12$ & $13.6\pm 2.0$ & $416\pm 16$ & RV+TTV \\ 
        \citet{Sinukoff2017} & $9.1\pm 1.2$ & $356\pm 12$ & $12.7\pm 2.7$ & $411\pm 18$ & RV \\ 
        \citet{Vanderburg2017} & $6.8\pm 0.7$ & $363\pm 7$ & $13.1\pm 1.5$ & $398\pm 9$ & RV$^\mathsection$ \\ 
        \citet{Bryant2022} & $6.8\pm 0.6$ & $364\pm 7$ & $14.2\pm 1.3$ & $399\pm 9$ & RV$^\S$ \\ \hline
        This work (see also Table~\ref{tab:trades_parameters}) & $9.0\pm 0.5$ & $374\pm 17$ & $15.5\pm 0.8$ & $447\pm 20$ & RV+TTV \rule{0pt}{14pt}\\ \hline
    \end{tabular}
    \tablefoot{$\dagger$: ``Detected weakly if at all'' \citep{Dai2015}. $\ddagger$: Upper limit at 95\% confidence \citep{Becker2015}. \valerio{$\mathsection$: The errors on $M_\mathrm{b}$ and $M_\mathrm{c}$ are dominated by the relative uncertainty on the stellar mass $M_\star$ (see Section~\ref{sec:discussion_conclusions}).} $\mathparagraph$: Photodynamical approach.}
    \label{tab:masses}
\end{table*}

Hundreds of multiple planetary systems have been detected with space-based missions such as Convection, Rotation and planetary Transits (CoRoT; \citealt{Auvergne2009}), Kepler/K2 \citep{Borucki2010,Ricker2015}, and the Transiting Exoplanet Survey Satellite (TESS; \citealt{Ricker2015}). These systems show great diversity in their architecture, with planets spanning the full range of planetary masses and being arranged in very different dynamical configurations. However, HJs have been known for a long time to be preferentially lonely when compared to other classes \citep{Latham2011,Huang2016,Steffen2012}, 
and almost all their known companions are long-period massive planets at $P\gtrsim 200$~d detected through either modulation or long-term trends in their RVs \citep{Knutson2014}. The rarity of HJ companions with inner or short-period orbits has been confirmed by a recent global analysis of TESS photometry by \citet{Hord2021}. There are only a handful of notable examples of HJs with nearby companions: WASP-47, the subject of this paper (see below), Kepler-730 \citep{Zhu2018,Canas2019}, TOI-1130 \citep{Huang2020}, WASP-132 \citep{hord2022}, and the most recently reported TOI-2000 \citep{Sha2021}. Together, these five planets represent less than $1\%$ of the known HJs\footnote{Source: NASA Exoplanet Archive. \url{https://exoplanetarchive.ipac.caltech.edu/}}. Notably, WASP-47 is unique among the mentioned systems for showing clear transit time variations (TTVs), allowing researchers to jointly exploit TTVs and RVs to measure the planetary masses and other orbital parameters in a much more reliable way than using the two techniques separately \citep{Malavolta2017}.

WASP-47 has been dubbed ``the gift that keeps on giving'' \citep{Kane2020} due to the many layers of scientific investigation it has stimulated. The first member of this planetary system, HJ WASP-47b\footnote{All the $M_\mathrm{p}$, $R_\mathrm{p}$, and $P$ values quoted in this paragraph are consistently borrowed from the analysis by \citet{Vanderburg2017}.} ($1.14\pm0.02$~$M_\mathrm{jup}$, $1.127\pm 0.013$~$R_\mathrm{jup}$, $P\simeq 4.159$~d) was discovered by \citet{Hellier2012} from ground-based photometric data (SuperWASP). A few years later, WASP-47 was observed by K2 during its Campaign 3, enabling the detection of two additional companions \citep{Becker2015}: WASP-47e  ($6.8\pm0.7$~$M_\oplus$, $1.81\pm 0.03$~$R_\oplus$, $P\simeq 0.789$~d), an inner super-Earth, and WASP-47d ($13.1\pm1.5$~$M_\oplus$, $3.58\pm 0.05$~$R_\oplus$, $P\simeq 9.031$~d), an outer Neptunian. Around the same time, the results from the first extensive RV campaign \citep{NeveuVanmalle2016} revealed the presence of a fourth non-transiting Jupiter-sized planet on a much longer orbit, WASP-47c  ($1.25\pm0.03$~$M_\mathrm{jup}$, $P\simeq 588$~d). Subsequent RV analysis with fresh data have further refined and/or constrained the planetary masses \citep{Dai2015,Sinukoff2017,Vanderburg2017}. The opportunity of combining TTV and RV analysis in a joint approach was exploited by \citet{Almenara2016} and \citet{Weiss2017}, the former within a \valerio{photodynamical framework.}\footnote{\valerio{We refer the reader to \citet{Almenara2015} for a complete description and discussion on the photodynamical approach.}} All the published measurements of planetary masses for WASP-47 are summarized in Table~\ref{tab:masses}. It is worth mentioning that the measurement of the spin-orbit angle obtained through the Rossiter-McLaughlin effect is consistent with an aligned configuration ($\lambda = 0^\circ\pm 24^\circ$) for WASP-47b  \citep{SanchisOjeda2015} and that a phase-curve re-analysis of K2 photometry constrained the WASP-47b albedo to values significantly lower than the average of HJs \citep{Kane2020}. Lastly, a very recent analysis by \cite{Bryant2022} focusing on the characterization of planet -e has made use of both TESS and ESPRESSO data for the first time.

Despite the extensive amount of work carried out on WASP-47 so far, a convincing and complete description of the evolutionary history of this system is still elusive. Our current understanding of this system will change once the James Webb Space Telescope (JWST) enables a detailed atmospheric characterization of WASP-47's transiting planets, especially of the small ones: -e and -d \citep{Bryant2022}. Unfortunately, a reasonably accurate prediction of the future transits of planet -d is difficult, as even the best available ephemerides based on a combination of K2 and RV data will have drifted by about one hour (1-$\sigma$) at epoch 2023.0. The photometric detection of WASP-47d is indeed beyond the reach of ground-based facilities, and TESS has been unable to significantly detect the only transit predicted to fall in Sector 42 (and will not re-observe it at least until the end of Cycle 6 in 2024). The need to recover a new set of reliable ephemerides is one of the initial motivations of our investigation. In addition, the mass measurements of planet -d and -e have historically been slightly inconsistent with each other due to a small statistical tension between the most precise estimate through a photodynamical approach and studies wholly based on RV data (Table~\ref{tab:masses}). Although the most recent ESPRESSO measurements \citep{Bryant2022} appear to have solved this tension for planet -d (but still not for -e), WASP-47 remains one of the very few systems for which a precise mass measurement can be achieved by either TTVs or RVs, potentially shedding some light on an old debate about a possible systematic discrepancy between the two techniques \citep{Mills2017,Petigura2018}. 

In this paper, we present new data from the  CHaracterising ExOPlanet Satellite (CHEOPS) and new ground-based photometric data for WASP-47b and WASP-47d, and we perform a global dynamical re-analysis of all the available data, including the latest TESS and ESPRESSO data sets. We focus particularly on the determination of the orbital and physical parameters of planet -d and -e in order to address the aforementioned issues. In Section~\ref{sec:observations}, we describe all the photometric and spectroscopic data analyzed in this work. In Section~\ref{sec:analysis}, we deal with the light curve fitting and the dynamical modeling. Finally, we present and discuss our results, including a comparison with the literature, in Section~\ref{sec:discussion_conclusions}.

\section{Observations}\label{sec:observations}

For our dynamical analysis, we gathered all the publicly available RV and photometric data of WASP-47. This included both public data (from K2, TESS, and several RV surveys) and new proprietary data from CHEOPS and ground-based telescopes, described below.
A log summarizing all the analyzed photometric data is reported in the Appendix. The time stamps associated with all the measurements were consistently converted to the $\textrm{BJD}_\textrm{TDB}$ standard, as prescribed by \citet{Eastman2010}.

\subsection{CHEOPS photometry}

Launched in 2019, CHEOPS \citep{Benz2021} is an ESA S-class mission currently carrying out its 3.5-year nominal observing program. The scientific instrument of CHEOPS is a 32-cm reflecting telescope designed to deliver defocused images for the performance of ultra high-precision photometric measurements of bright stars. The high-performance capabilities of CHEOPS when working with transit timing in particular have been demonstrated by \citet{Borsato2021b}.

\begin{figure*}
    \centering
    \includegraphics[width=0.9\columnwidth]{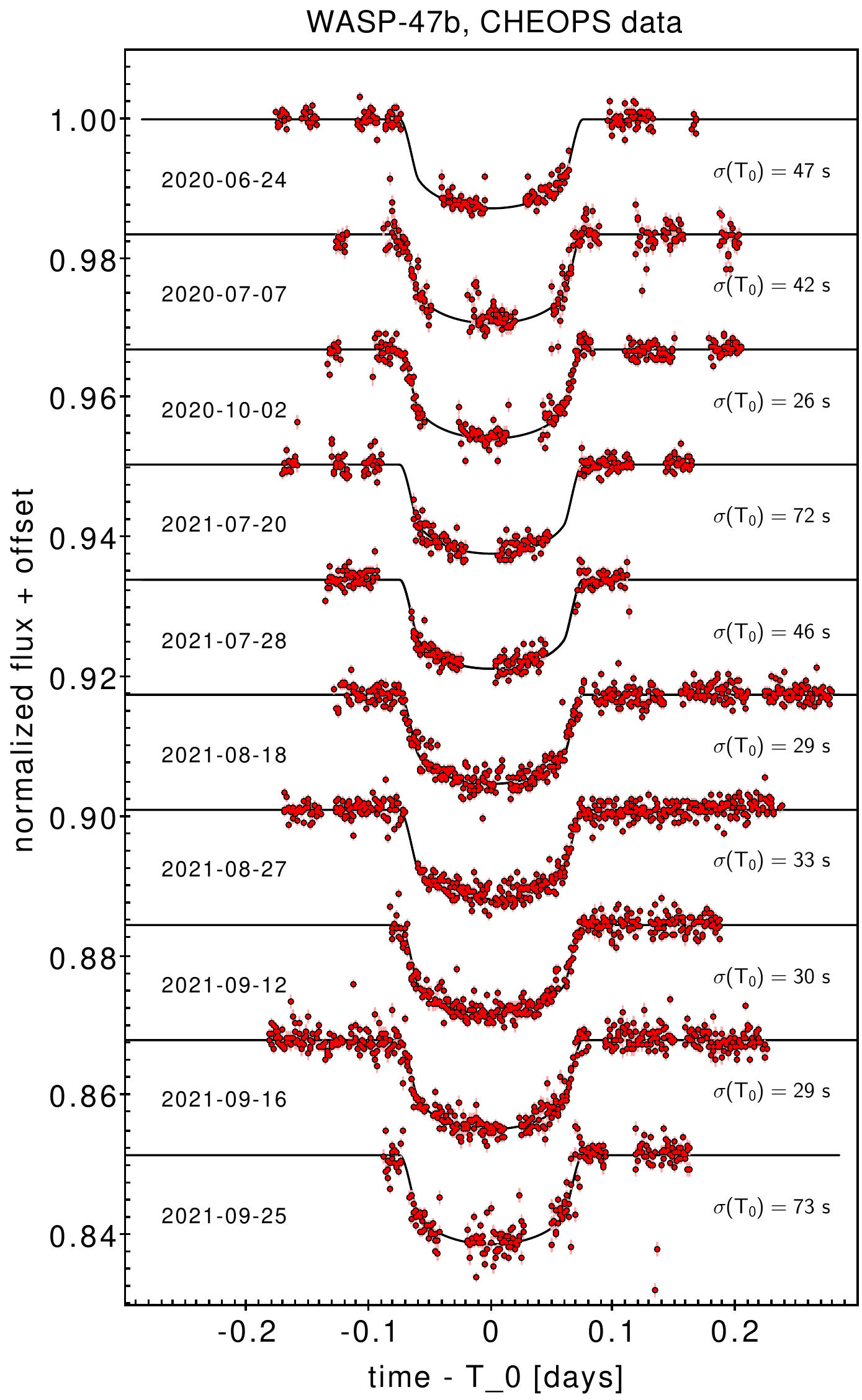}\hspace{0.05\columnwidth}
    \includegraphics[width=0.9\columnwidth]{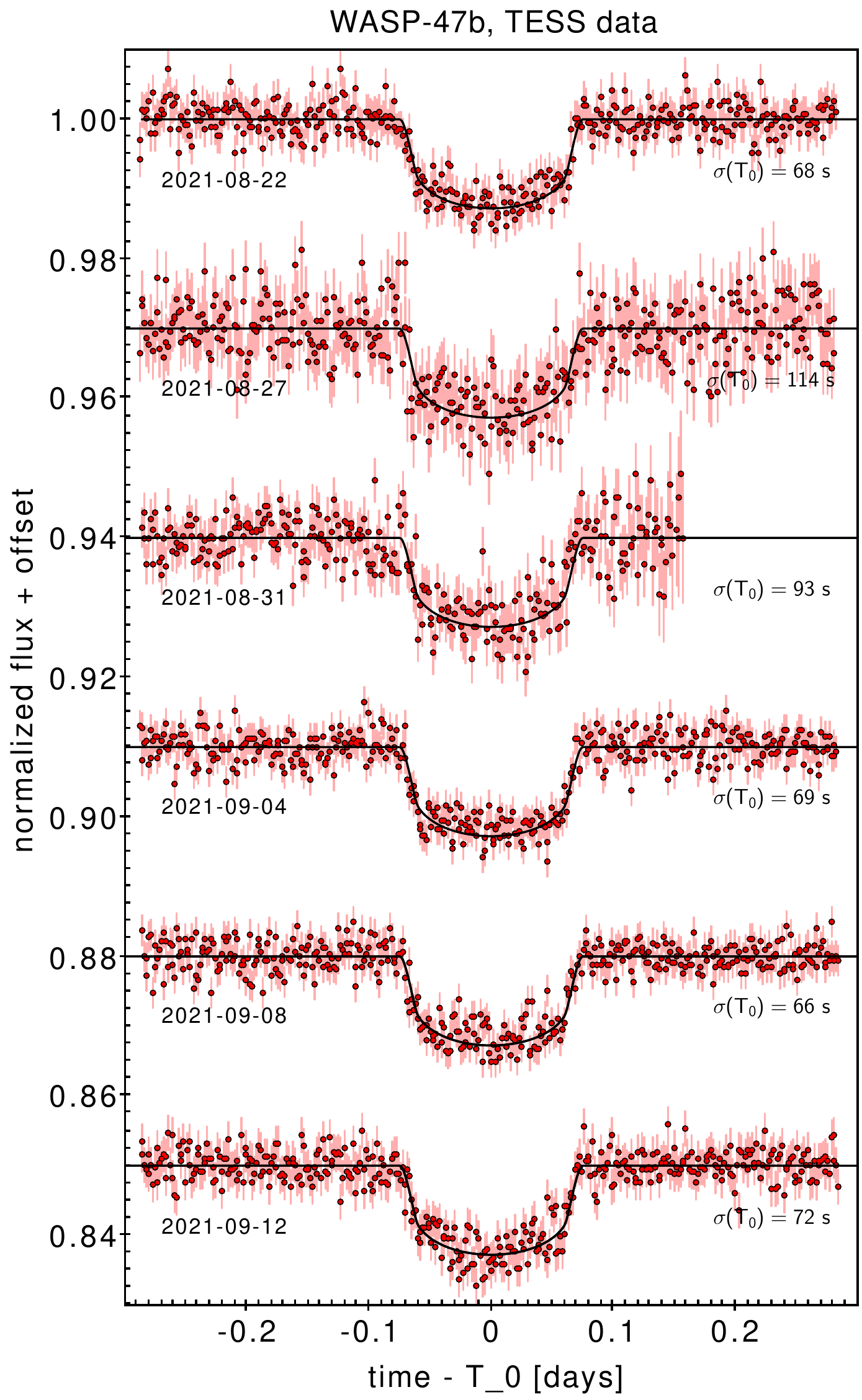}\\
    \caption{\valerio{Space-based} light curves of WASP-47b from CHEOPS  and TESS  analyzed for the present work after de-trending. For each light curve, the corresponding label shows the acquisition date in UT and the timing error $\sigma_{T_0}$ (in seconds). Arbitrary vertical offsets of 0.0165 and 0.03 were added, respectively, to both sets for visualization purposes.}
    \label{fig:lcsb}
\end{figure*}

Eleven visits of WASP-47 were scheduled with CHEOPS in 2020 and 2021 within the Guaranteed Time Observations (GTO) programs \texttt{CH\_PR100017} and \texttt{CH\_PR100025}  (see Table~\ref{tab:datasets} for a list of the CHEOPS data sets). Ten visits were centered on the transits of WASP-47b (Table~\ref{tab:tzero_b_space}), and one was centered on WASP-47d (Table~\ref{tab:tzero_d}). All the light curves were extracted from the raw satellite data by the official CHEOPS Data Reduction Pipeline v13.1 (DRP; \citealt{Hoyer2020}), and their plots are shown in the left panel of Fig.~\ref{fig:lcsb} (planet -b) and in Fig.~\ref{fig:lcsd} (planet -d). The exposure time was set to 60~s and the minimum efficiency to 50\%, resulting in some of the light curves showing noticeable gaps every 98.77~min (the orbital period of CHEOPS) due to 
the spacecraft crossings of the South Atlantic Anomaly as well as Earth-related constraints (blockage of the line-of-sight by the Earth and high levels of stray light).

\begin{figure*}
    \centering
    \includegraphics[width=0.9\columnwidth]{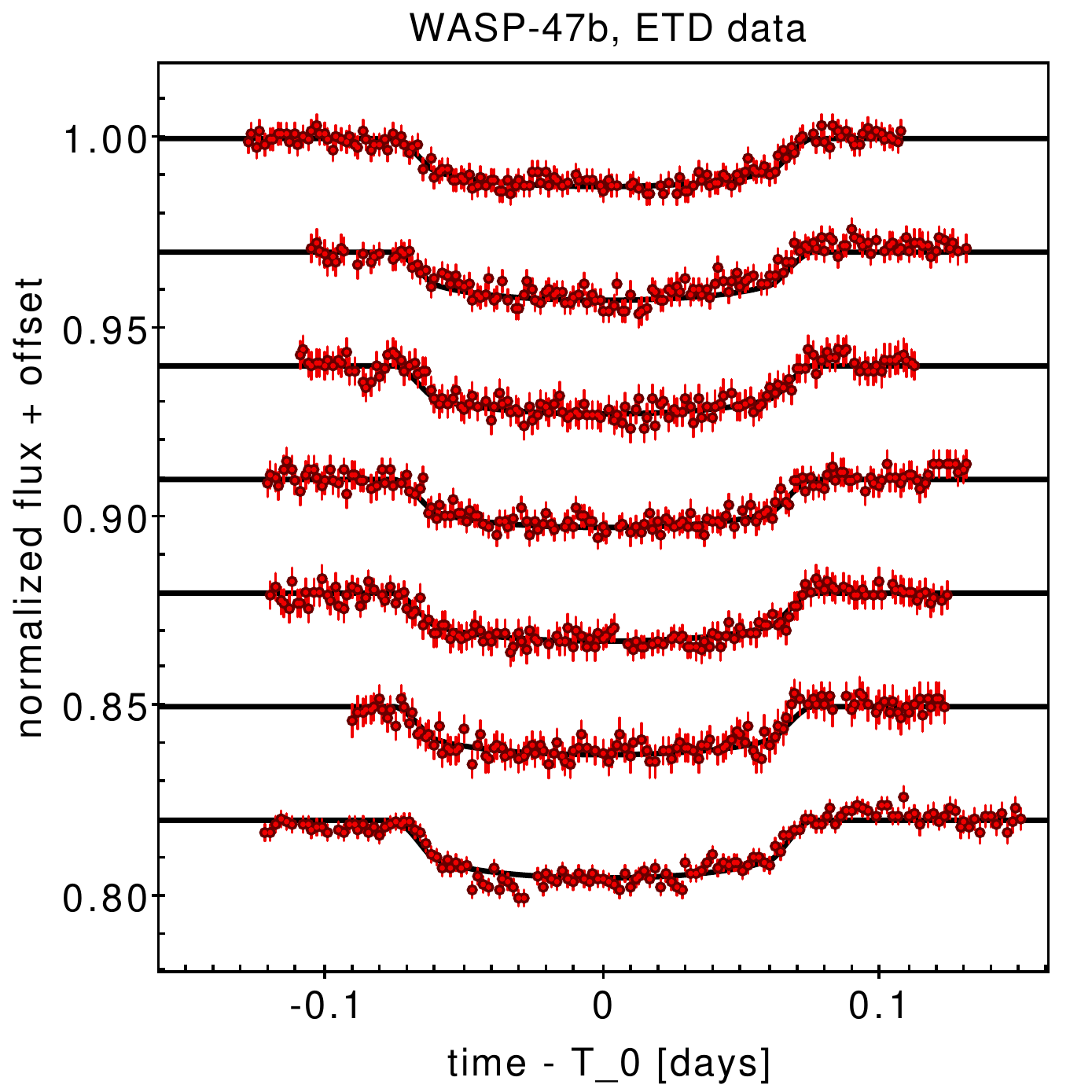}\hspace{0.05\columnwidth}
    \includegraphics[width=0.9\columnwidth]{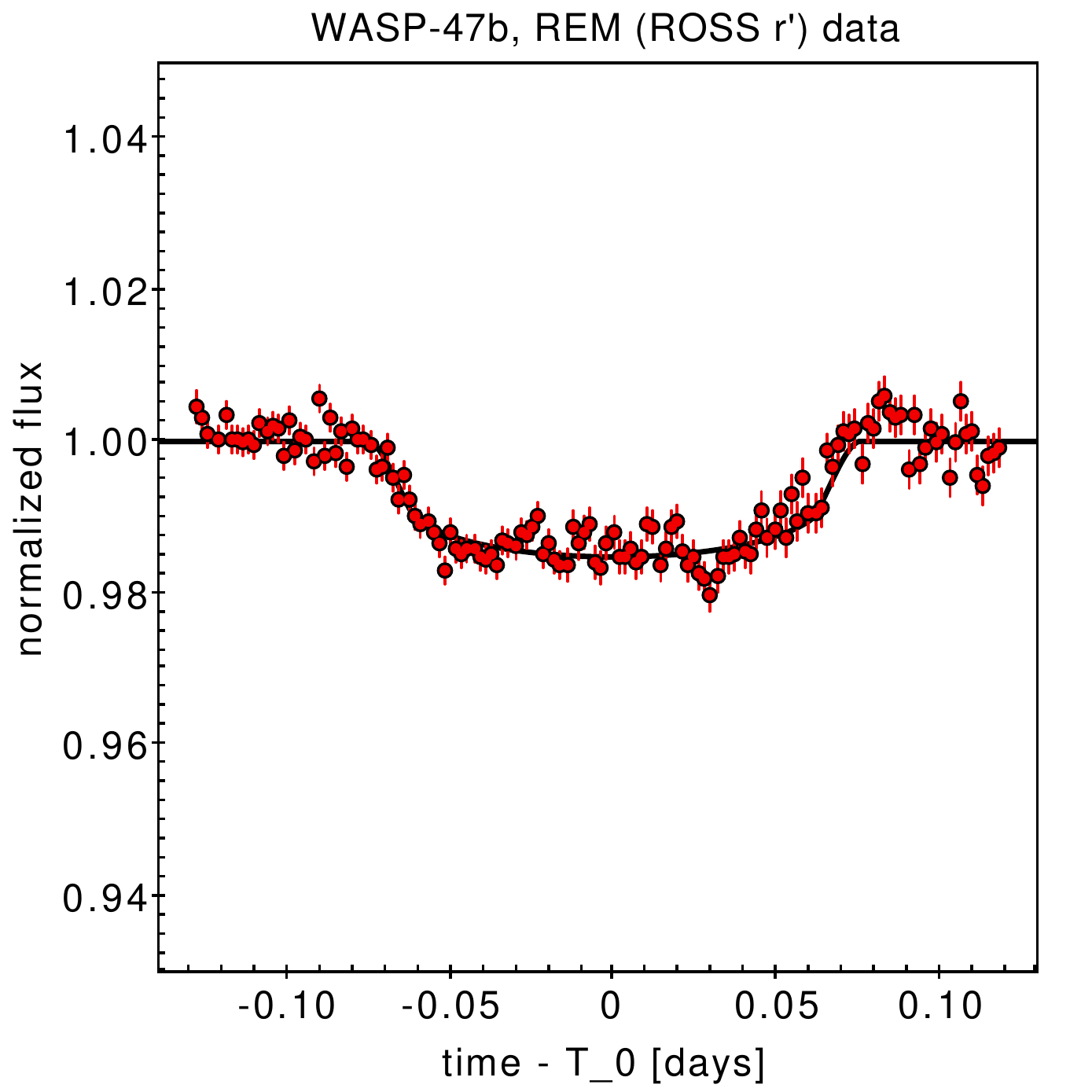}\\
    \caption{\valerio{Ground-based light curves of WASP-47b from ETD and REM analyzed for the present work. An arbitrary vertical offset of 0.03 was added to the ETD light curves for visualization purposes. The light curves are sorted in chronological order (see~Table~\ref{tab:tzero_b_ground}) increasing towards bottom.}}
    \label{fig:lcsb_ground}
\end{figure*}

\begin{figure}
    \centering
    \includegraphics[width=0.9\columnwidth]{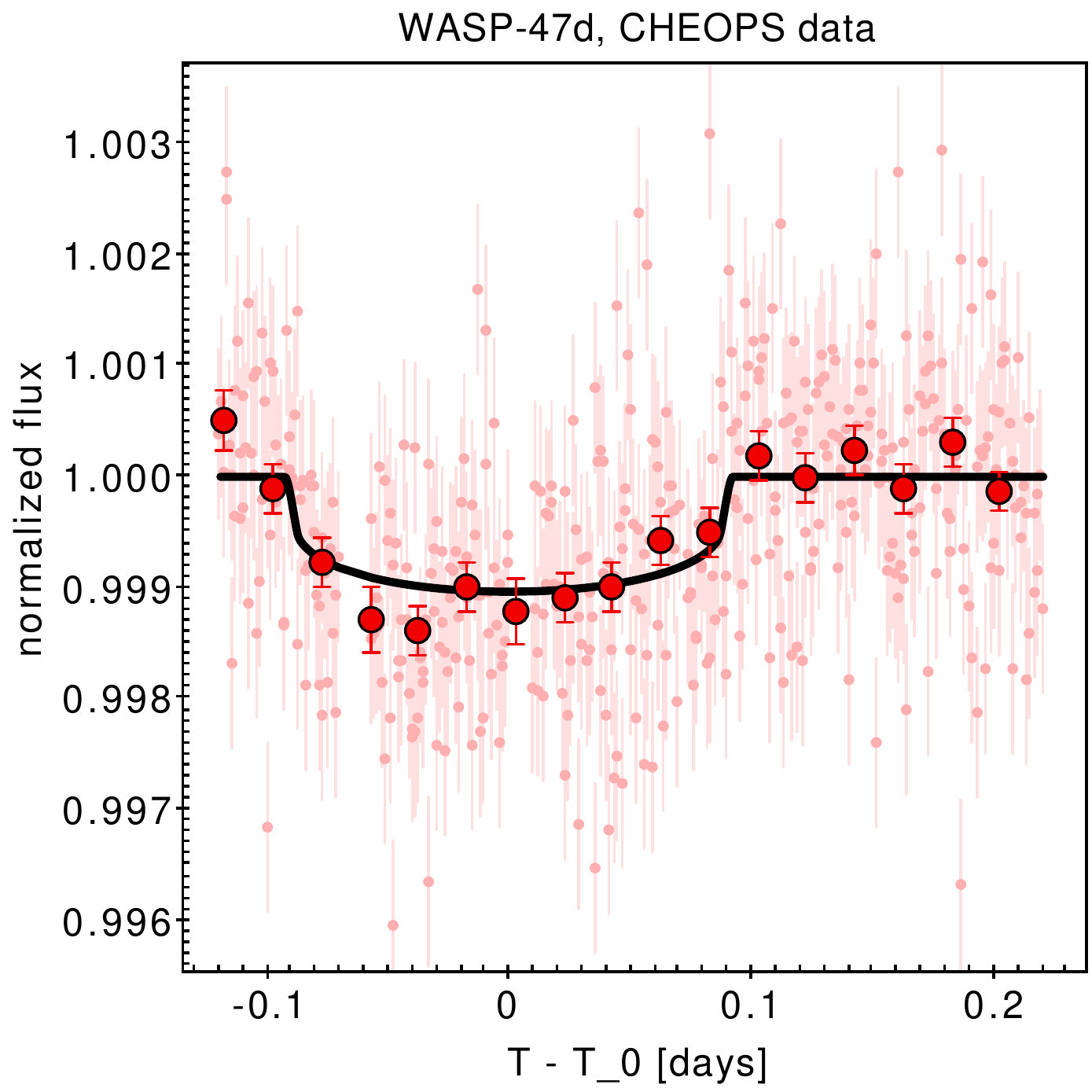}
    \caption{\valerio{CHEOPS light curve of WASP-47d after de-trending. The pink data points are at the original sampling cadence (60~s), and the red circles were averaged over 20-min intervals.}}
    \label{fig:lcsd}
\end{figure}

\subsection{K2 and TESS photometry}

During its Campaign 3, K2 monitored WASP-47 in short-cadence mode (net cadence: 58.3~s)  for 69 nearly uninterrupted days, from November 14, 2014, to January 23, 2015. These data sets were used to study the transits of planets -d and -e for the first time \citep{Becker2015}. We also use their light curves in our analysis, as they are already corrected for jitter-induced systematic errors by fitting splines as a function of spatial drift, following the procedure by \citet{Vanderburg2014}. Overall, 108 transit light curves from K2 were ingested in our analysis: 16 of planet -b, seven of planet -d, and 85 of planet -e. These are summarized in Table~\ref{tab:tzero_b_space}, \ref{tab:tzero_d}, and \ref{tab:tzero_e1}-\ref{tab:tzero_e2}, respectively.

Almost seven years later, WASP-47 was observed by TESS for the first time (and the only time so far) in Sector 42 (camera 1, CCD 4), from August 20 to September 16, 2021. Being a known planet host, it was designated as a 120-s cadence target. \valerio{The resulting data set is the same investigated by \citet{Bryant2022}, who used of the Pre-search Data Conditioned Simple Aperture Photometry (PDCSAP)\ light curve made available by the Science Processing Operations Center (SPOC) pipeline \citep{Jenkins2016}. Unfortunately, the PDCSAP light curve is missing a large one-week section, including two transits of WASP-47b (on August 27 and 31), due to a ``scattered light'' quality flag. 
Unlike \citet{Bryant2022}, we based our analysis on Simple Aperture Photometry (SAP) data, where the two missing transits are preserved and do not appear to be impacted by a significant amount of systematic noise, as is demonstrated in Section~\ref{sec:analysis_lc}. While the SAP algorithm does not correct the light curves for photometric dilution due to contaminating sources, we emphasize that no Gaia~DR3 source (limiting magnitude $G\simeq 21$) is present at all within a radius of $30''$ from WASP-47, and the brightest one within $60''$ has $G=17.50$, implying a negligible dilution factor. We also note that a constant dilution factor does not have any systematic effect on the measurement of transit times because it does not alter the symmetry of the transit signal.} 

Overall, six TESS light curves of planet -b (Table~\ref{tab:tzero_b_space}; right panel of Fig.~\ref{fig:lcsb}) were ingested.
As already noted by \citet{Bryant2022}, individual transits from planet -d and -e are hidden in the TESS photometric scatter, and their detection is at most marginal. Our preliminary fitting tests confirmed this, and we did not attempt to retrieve their timings.
The next visit to WASP-47  by TESS is not expected until at least the end of Cycle 6 (October 2024).  

\subsection{Ground-based photometry}

One transit of WASP-47b was observed on September 24, 2021, by the 60-cm  Rapid-Eye Mount telescope (REM; \citealt{Zerbi2001}) through the ROS2 instrument \citep{Molinari2014}, a charge-coupled device (CCD) camera able to observe the same field of view ($9.1'\times 9.1'$) simultaneously in four photometric bands (Sloan $g'$, $r'$, $i'$,  and $z'$). These observations (PI: Nascimbeni) were carried out as part of the TASTE program, a ground-based multi-site long-term TTV campaign to monitor transiting planets \citep{Nascimbeni2011}. Two bright reference stars (TYC~5805-338-1 and TYC~5805-739-1) were always imaged along with the target, allowing us to perform differential photometry. In addition, in order to mitigate systematic errors from guiding drifts, pixel-to-pixel non-homogeneity, and flat-field errors, a generous amount of defocus was applied ($\textrm{FWHM}\gtrsim 10$~pix). 

The four transit light curves from the observation were extracted by running STARSKY \citep{Nascimbeni2013}, the photometric pipeline developed for TASTE. The STARSKY pipeline automatically selects the best combination of photometric aperture radii and weighting scheme to minimize the final out-of-transit scatter. \valerio{No linear detrending against the external parameters automatically tested by STARSKY proved to be effective (i.e., linear baseline, airmass, $x$-$y$ position, background level, stellar FWHM).} Guiding drifts unfortunately turned out to be much larger than anticipated, totaling more than 30 pixels throughout the whole series. The resulting systematic errors impacted the $g'$, $i'$,  and $z'$ data to an unacceptable extent, while the $r'$ light curve was mostly spared, thanks to its much better flat-field correction and its structure being more homogeneous at small spatial scales. For this reason, we only considered the $r'$ light curve (right panel of Fig.~\ref{fig:lcsb_ground}) for our subsequent analysis.

Seven additional transit light curves of WASP-47b (of six distinct transit events) were collected from 2020 to 2021 by Y.~Jongen and A.~W\"unsche and are available on the public archive of the Exoplanet Transit Database (ETD; \citealt{Poddany2010}). They were gathered through an $R$ filter with a PlaneWave CDK 17" telescope coupled with a Moravian~G4~16k CCD camera at the Deep Sky Chile facility (Jongen) and with an unfiltered PlaneWave CDK 20" telescope coupled with the same camera at the El Sauce Observatory in Chile (W\"unsche). The exposure time was set to 120~s by both observers.
After visually checking their quality and converting the time stamps to $\textrm{BJD}_\textrm{TDB}$, the transit light curves were included in our analysis as well. These light curves are detailed in Table~\ref{tab:tzero_b_ground} \valerio{and plotted in the left panel of Fig.~\ref{fig:lcsb_ground}.}

\subsection{Radial velocities}

We collected all the available high-precision RV data for our analysis from the scientific literature:
\begin{itemize}
\item 19 from ESPRESSO \citep{Bryant2022}, 
\item 47~from~HIRES \citep{Sinukoff2017}, 
\item 69~from~HARPS-N \citep{Vanderburg2017}, 
\item 26~from~PFS \citep{Dai2015}, 
\item 51~from~CORALIE \citep{Hellier2012, NeveuVanmalle2016}, 
\end{itemize}
for a total of 212 independent data points spanning nine years (2010-2019). 

\section{Data analysis}\label{sec:analysis}

\begin{table}[!t]
    \centering\renewcommand{\arraystretch}{1.2}
    \caption{Stellar and limb darkening parameters adopted in the transit fitting process (Section~\ref{sec:analysis_lc}) and subsequent dynamical analysis (Section~\ref{sec:analysis_dynamical}).}
    \begin{tabular}{ccc} 
    \multicolumn{3}{c}{Stellar parameters} \\
    \hline\hline
    \multicolumn{2}{l}{Mass ($M_{\odot}$)} & $1.058 \pm 0.047$ \\
    \multicolumn{2}{l}{Radius ($R_{\odot}$)} & $1.156 \pm 0.009$ \\ \hline
    \\
    \\
    \multicolumn{3}{c}{Limb darkening parameters} \\
    \hline\hline
    Instrument & $u_1$ & $u_2$ \\
    \hline
    CHEOPS  & $0.571 \pm 0.020$ & $0.096 \pm 0.044$\\
    REM ($r$) & $0.558 \pm 0.023$ & $0.112 \pm 0.054$\\
    TESS    & $0.444 \pm 0.016$ & $0.120 \pm 0.038$\\
    K2      & $0.562 \pm 0.020$ & $0.102 \pm 0.044$\\
    ETD     & $0.558 \pm 0.023$ & $0.112 \pm 0.054$\\
    \hline
    \end{tabular}
    \tablefoot{Both sets of parameters were derived as described in Section~\ref{sec:analysis_lc}.}
    \label{tab:star}
\end{table}

\subsection{Light curve fitting}\label{sec:lc_fit} \label{sec:analysis_lc}

All the photometric data described in the previous section were homogeneously (re-)analyzed through the \texttt{PyORBIT} \citep{Malavolta2016,Malavolta2018} code. The CHEOPS light curves also went  through an additional detrending stage
using \texttt{cheope}\footnote{\url{https://github.com/tiziano1590/cheops_analysis-package}}, an optimized \texttt{python} tool (of which \texttt{pycheops} is the back-end; \citealt{Maxted2021}) 
specifically developed to filter, correct, and fit CHEOPS data.
The detrending model of \texttt{cheope} is defined as a linear combination of terms, including a quadratic baseline  $f_0 + \mathrm{d}f/\mathrm{d}t + \mathrm{d}^2f/\mathrm{d}t^2$, the first and second order derivative of the centroid offset in $x$ and $y$ pixel coordinates ($\mathrm{d}f/\mathrm{d}x$, $\mathrm{d}^2f/\mathrm{d}x^2$, $\mathrm{d}f/\mathrm{d}y$,  $\mathrm{d}^2f/\mathrm{d}y^2$), background ($\mathrm{d}f/\mathrm{d}b$), contamination ($\mathrm{d}f/\mathrm{d}\textrm{contam}$), and the first three harmonics of the spacecraft roll angle (in $\cos\phi$ and $\sin\phi$). There is also an additional term called ``glint'' that empirically models the internal reflections as a spline-based smooth function \citep{Borsato2021b}. The combination of terms to be adopted as the best detrending model was selected automatically by \texttt{cheope} according to its Bayes factor, first through an \texttt{lmfit} optimization \citep{Newville2014} to find a reasonable starting point and then with an \texttt{emcee} fit \citep{Foreman2013}.

For both CHEOPS and all the other data sets, the stellar activity signal and any residual instrumental trend were subtracted using \texttt{w\=otan} \citep{Hippke2019} after masking the transits of planets -b, -d, and -e  to get the normalized light curves. The filter adopted was a Tukey's bi-weight with a window filter duration set equal to the transit duration plus its 1-$\sigma$ uncertainty.
All the detrended transits from ground- and space-based telescopes are shown in Figs.~\ref{fig:lcsb}, \ref{fig:lcsb_ground}, and \ref{fig:lcsd}. 
As for the TESS Sector 42 light curves, we focused our analysis on the extraction of six transits of planet -b only, after masking the transits of planets -d and -e, which were only marginally detectable in this data set and hence not usable for an individual $T_0$ fit.

In order to extract the central transit times $T_0$, we applied \texttt{PyORBIT} independently on each of the five data sets (i.e., CHEOPS, TESS, K2, REM, and ETD), performing a modeling of each light curve using \texttt{PyDE}+\texttt{emcee} \citep{Parviainen2016,Foreman2013}. We assumed Keplerian orbits for all three fitted planets, -b, -e, and -d. We fixed the $a/R_\star$ parameter of each planet at the values found on the highest S/N data set (K2; \citealt{Vanderburg2017}), leaving the individual $T_0$ values, the planetary radius $R_\mathrm{p}/R_\star$, and the orbital inclination $i$  for each planet as the only free parameters in order to take into account changes in the impact parameter as a function of time or different transit depths through different instruments and filters. 
 The stellar parameters were derived by the CHEOPS Stellar Characterization team following the procedure described by \citet{Borsato2021a}, Section~3.2.1. The two parameters that are crucial for our dynamical analysis, the stellar radius ($R_\odot$) and mass ($M_\odot$), are listed in Table~\ref{tab:star}. We adopted a quadratic law $u_1$, $u_2$ for the limb darkening coefficients, but we re-parameterized them as  fitting parameters $q_1$ and $q_2$ , as done by \citet{Kipping2013}. The priors set on the coefficients $u_1$ and $u_2$ are reported in Table~\ref{tab:star}. We assumed a Gaussian prior distribution over the coefficients for all the instruments using a bi-linear interpolation of the limb-darkening profile defined in \citet{Claret2017,Claret2021}. 
 
We used \texttt{PyDE} to find the best starting point for the MCMC sampler. For each data set, \texttt{PyDE} used a population size of $4\times N_{par}$, where $N_{par}$ is the number of free parameters, and let it evolve for $64\,000$ generations, and for the Bayesian sampler, we set 100 chains for $150\,000$ steps, discarding the first $40\,000$ by adopting a thinning factor of 100. All the final $T_0$ values of this analysis are summarized in Tables~\ref{tab:tzero_b_space} to \ref{tab:tzero_e2} of Appendix \ref{sec:appendix}, while the physical best-fit values of the common parameters among individual data sets 
($R_\mathrm{p}$, $i$)
are shown in Table~\ref{tab:planet}. \valerio{As anticipated in Section~\ref{sec:observations}, we directly compared the transit timings we extracted from TESS SAP photometry with those obtained from the PDCSAP light curves by \citet{Bryant2022} and listed in their Table~4 to check for any systematic offset. The average $\Delta T_0 = T_{0,\textrm{PDCSAP}}-T_{0,\textrm{SAP}}$ evaluated over the four transits in common yielded 14~s (with average error bars of $\sim 66$~s), and no pair disagreed by more than 0.5~$\sigma$. In other words, timings from SAP and PDCSAP are in perfect statistical agreement.
}

\begin{table}
    \centering\renewcommand{\arraystretch}{1.1}
    \caption{Best-fit values of planetary radius $R_\mathrm{p}$ (derived from $R_\mathrm{p}/R_\star$) and orbital inclination $i$. All these parameters were fitted as common parameters among individual data sets.}
    \begin{tabular}{cccc} \hline\hline
    Parameter & WASP-47b & WASP-47d & WASP-47e \\
    \hline
    \multicolumn{1}{c}{CHEOPS} \rule{0pt}{12pt} & & &  \\ 
    $R_\mathrm{p}\, (R_{\oplus})$ & $12.72 \pm 0.12$ & $\phantom{0}3.88 \pm0.20$ & -- \\     
    $i$\,($^\circ$) & $89.21 \pm 0.51$ & $89.23 \pm 0.51$& -- \\
    \multicolumn{1}{c}{TESS} \rule{0pt}{12pt} & & &  \\
    $R_\mathrm{p}\, (R_{\oplus})$ & $12.61 \pm 0.17$ & -- & -- \\     
    $i$\,($^\circ$) & $88.6 \pm 0.6$ & -- & -- \\
    \multicolumn{1}{c}{K2} \rule{0pt}{12pt} & & &  \\
    $R_\mathrm{p}\, (R_{\oplus})$ & $12.86 \pm 0.10$ & $\phantom{0}3.65 \pm 0.03$& $\phantom{0}1.83 \pm 0.02$ \\     
    $i$\,($^\circ$) & $89.00 \pm 0.17$ & $89.23 \pm 0.10$& $86.73 \pm 1.76$ \\
    \multicolumn{1}{c}{ETD} \rule{0pt}{12pt} & & &  \\
    $R_\mathrm{p}\, (R_{\oplus})$ & $12.95 \pm 0.17$ & -- & -- \\     
    $i$\,($^\circ$) & $89.06 \pm 0.58$ & -- & -- \\
    \multicolumn{1}{c}{REM} \rule{0pt}{12pt} & & &  \\
    $R_\mathrm{p}\, (R_{\oplus})$ & $13.64 \pm 0.28$ & -- & -- \\     
    $i$\,($^\circ$) & $87.86 \pm 0.45$ & -- & -- \\
    \hline
    \end{tabular}
    \tablefoot{See Section~\ref{sec:analysis_lc} for details about the transit fit.}
    \label{tab:planet}
\end{table}

\begin{figure*}
    \centering
    \includegraphics[width=1.8\columnwidth]{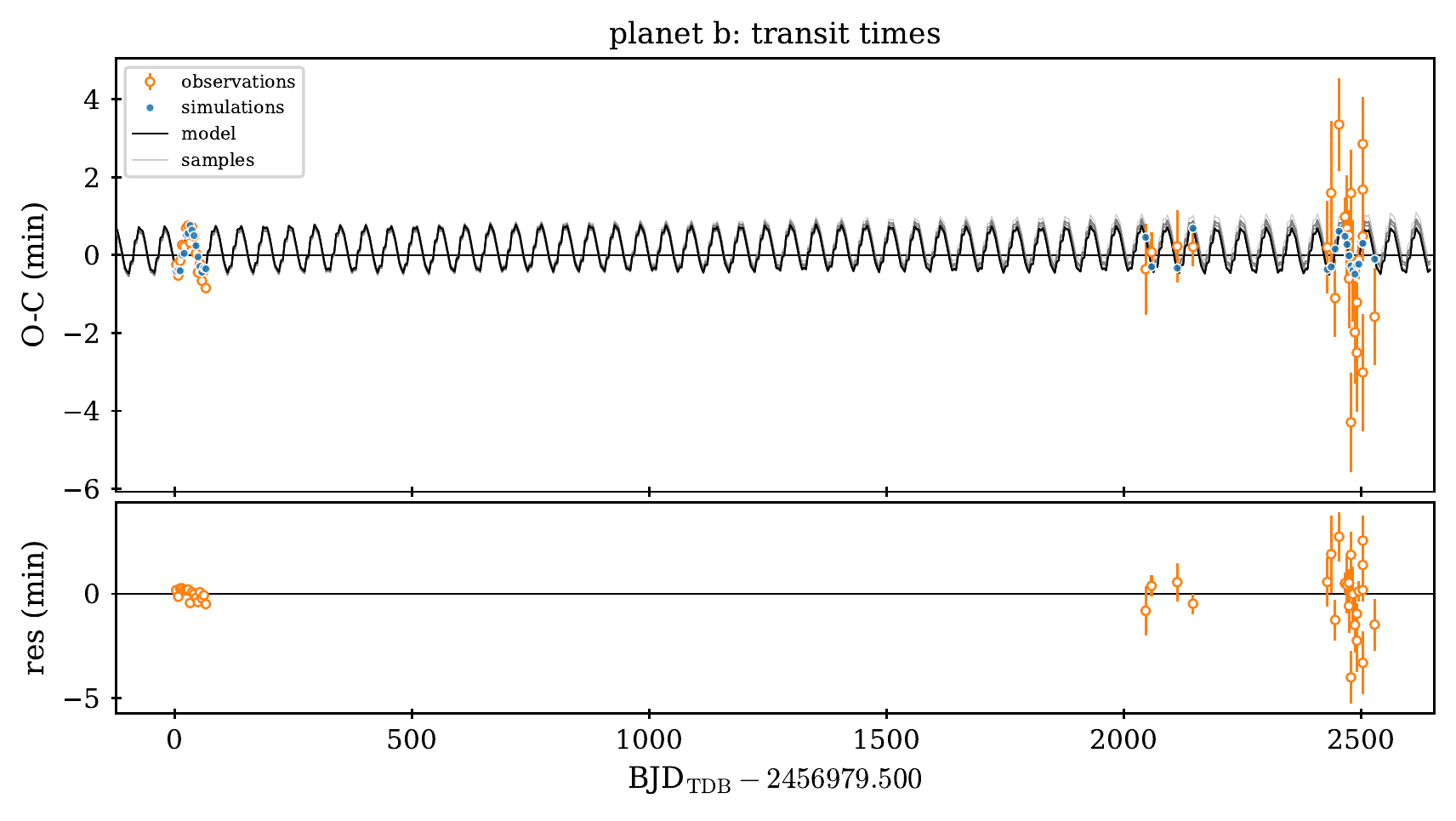}
    \caption{\valerio{Transit times of WASP-47b. Upper panel: Observed - Calculated $(O-C)$ diagram. The $O-C$ was calculated by subtracting the predicted $T_0$ calculated from the linear ephemeris to the observed transit times. The $O-C$ values computed from the observed $T_0$s are plotted as open orange  circles, while the $O-C$ of the TRADES-simulated $T_0$s from the best-fit model are plotted as blue circles. Samples drawn from the posterior distribution within HDI are shown as grey lines.
    Lower panel: Residuals computed as the difference between observed and simulated $T_0$s.}}
    \label{fig:trades_oc_b}
\end{figure*}

\begin{figure*}
    \centering
    \includegraphics[width=1.8\columnwidth]{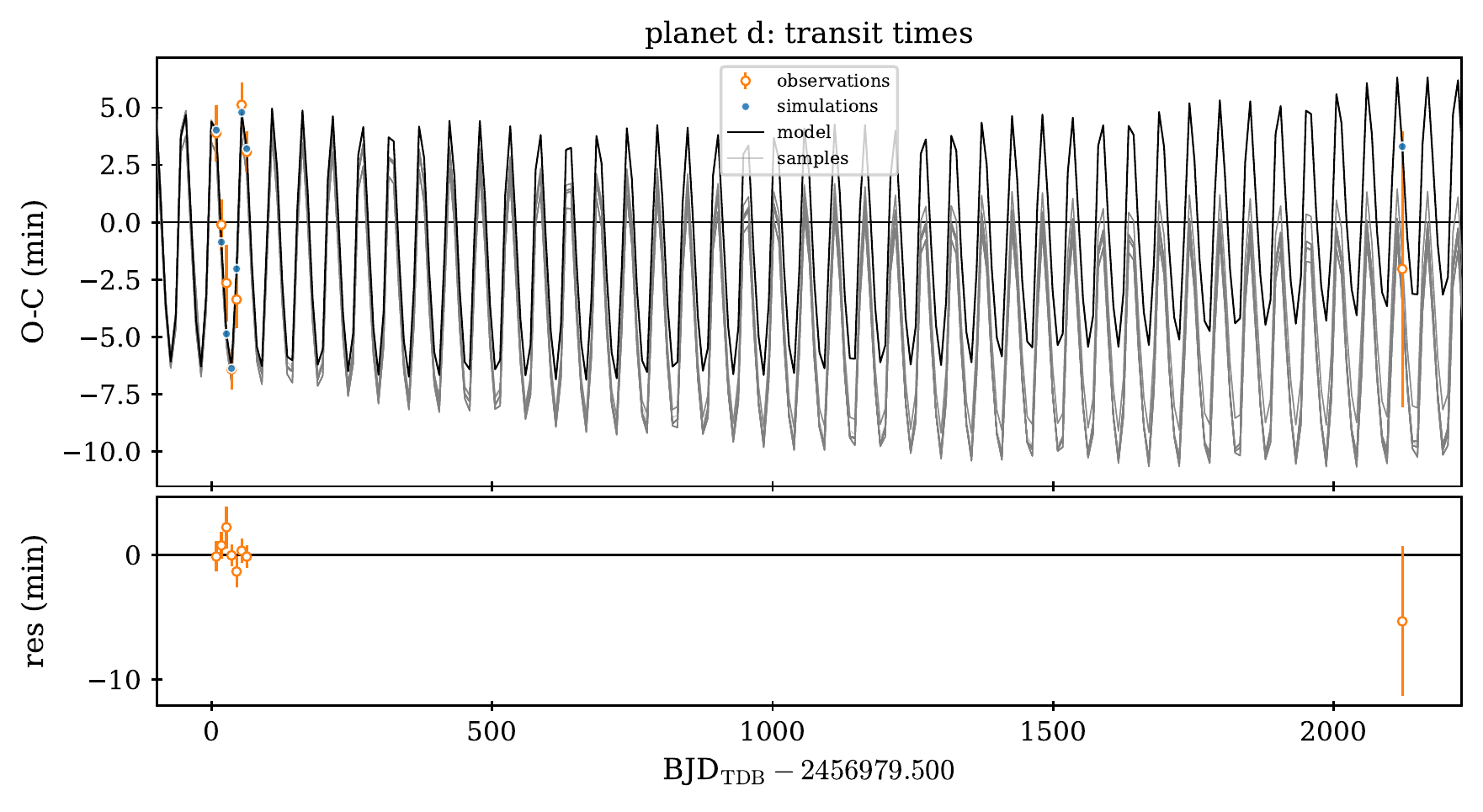}
    \caption{\valerio{Transit times of WASP-47d. Upper panel: Observed - Calculated $(O-C)$ diagram. The $O-C$ was calculated by subtracting the predicted $T_0$ calculated from the linear ephemeris to the observed transit times. The $O-C$ values computed from the observed $T_0$s are plotted as open orange circles, while the $O-C$ of the TRADES-simulated $T_0$s from the best-fit model are plotted as blue circles. Samples drawn from the posterior distribution within HDI are shown as grey lines.
    Lower panel: Residuals computed as the difference between observed and simulated $T_0$s.} }
    \label{fig:trades_oc_d}
\end{figure*}

\subsection{Dynamical modeling}\label{sec:analysis_dynamical}

We employed the TRADES dynamical integrator \citep{Borsato2014}, which has already been successfully applied to Kepler/K2 \citep{Borsato2019} and CHEOPS data \citep{Borsato2021b} as well as to simulate the Ariel TTV science case \citep{Borsato2021a}.
The TRADES dynamical integrator allowed us to fit transit times ($T_0$) and RVs simultaneously
during the integration of the orbits.
We fitted the mass ratios ($M_\mathrm{p}/M_\star$),
the periods ($P$), 
and the mean longitude ($\lambda$) for all the planets.
The eccentricity ($e$) and the argument of periastron ($\omega$) 
were fitted as ($\sqrt{e}\cos\omega, \sqrt{e}\sin\omega$) for planets -b, -c, and -d,
while we used the (initial) circular orbit ($e = 0$, $\omega=90^\circ$) for planet -e due to its extremely short tidal circularization timescale \citep{Vanderburg2017}.
We adopted our initial values for $M_b$, $M_c$, $M_d$, $M_e$, $e$ and $\omega$ from \citet{Almenara2016}.
We fixed the longitude of the ascending node ($\Omega$) of -b, -c, and -e to $180^\circ$.
For the orbital inclination $i$, we took the weighted average of the fitted values
from Section~\ref{sec:analysis_lc} and Table~\ref{tab:planet}
(for instance, $i_\mathrm{b} = 88.88\pm0.14^\circ$). 
We found that by assuming $i_\mathrm{d} < 90^\circ$,
one of the most recent transits of planet -b was missed by our integration
(i.e., the impact parameter for that transit was greater than one).
\citet{Almenara2016} reported a best-fit inclination higher than $90^\circ$,
but the authors specified that the supplementary angle is equally probable.
If both planets -b and -d transit in the same stellar hemisphere 
(with respect to the observer's line of sight), then
their relative distance would be shorter with respect to the case with planet -d on the opposite hemisphere.
The mutual gravitational interaction would of course be much stronger in the former case,
moving planet -b on some occasions to a non-transiting configuration that is not observed in our data.
Therefore, we assumed that planet -d transits the opposite hemisphere with respect to -b ($i_\mathrm{b} < 90^\circ$)
and adopted the value $180^\circ - i_\mathrm{d}$ as the initial parameter for \texttt{TRADES}.
We also fitted the $i$ and $\Omega$ of -d, and we set $i_\mathrm{c}$ to $90^\circ$.
All the orbital parameters are astrocentric and
defined at the reference time $2456979.5\, \mathrm{BJD_{TDB}}$, 
the same as in \citet{Almenara2016}.
The initial periods were set to the linear ephemeris fitted to $T_0$s determined in Section~\ref{sec:analysis_lc}.
For each RV data set, we fitted a $\log_2$-based jitter and an RV offset ($\gamma$).
However, we found that changing the starting point of the jitter
(e.g., $\log_{2}$ of 0.5 and $<1\times10^{-6}\,\mathrm{ms^{-1}}$)
did not affect the final distribution of the chains and, therefore, did not change the best-fit solution.
We split the CORALIE data set into three subsets, as done in \citet{Almenara2016}, for a total of seven RV data sets.
In all, we fitted 34 parameters, all with uniform-uninformative priors in the parameter space.

\begin{figure}
    \centering
    \includegraphics[width=0.99\columnwidth]{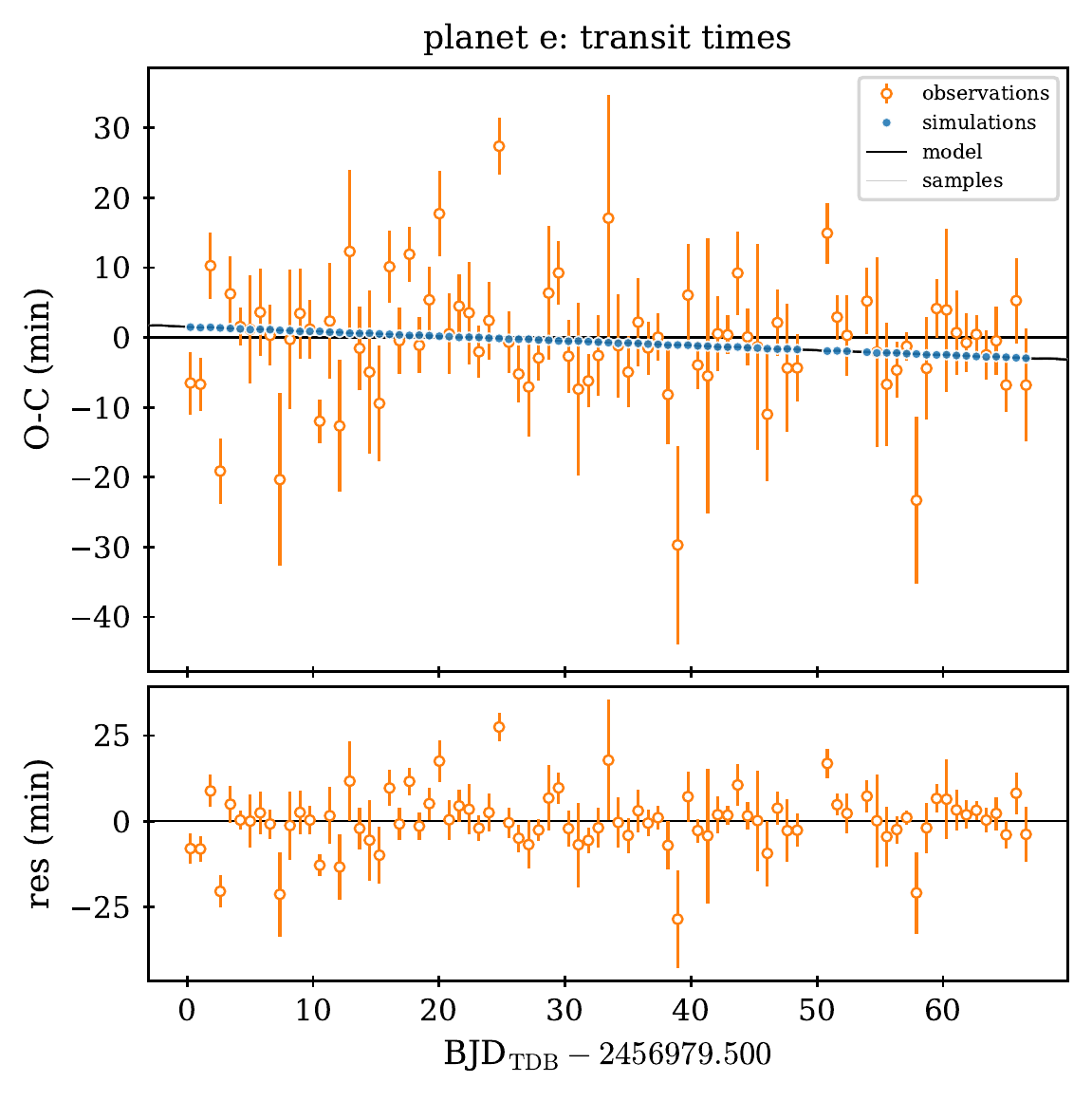}
    \caption{Transit times of WASP-47e. Elements are the same as in Figs.~\ref{fig:trades_oc_b} and \ref{fig:trades_oc_d}, but for planet -e.}
    \label{fig:trades_oc_e}
\end{figure}

\begin{figure*}
    \centering
    \includegraphics[width=1.99\columnwidth]{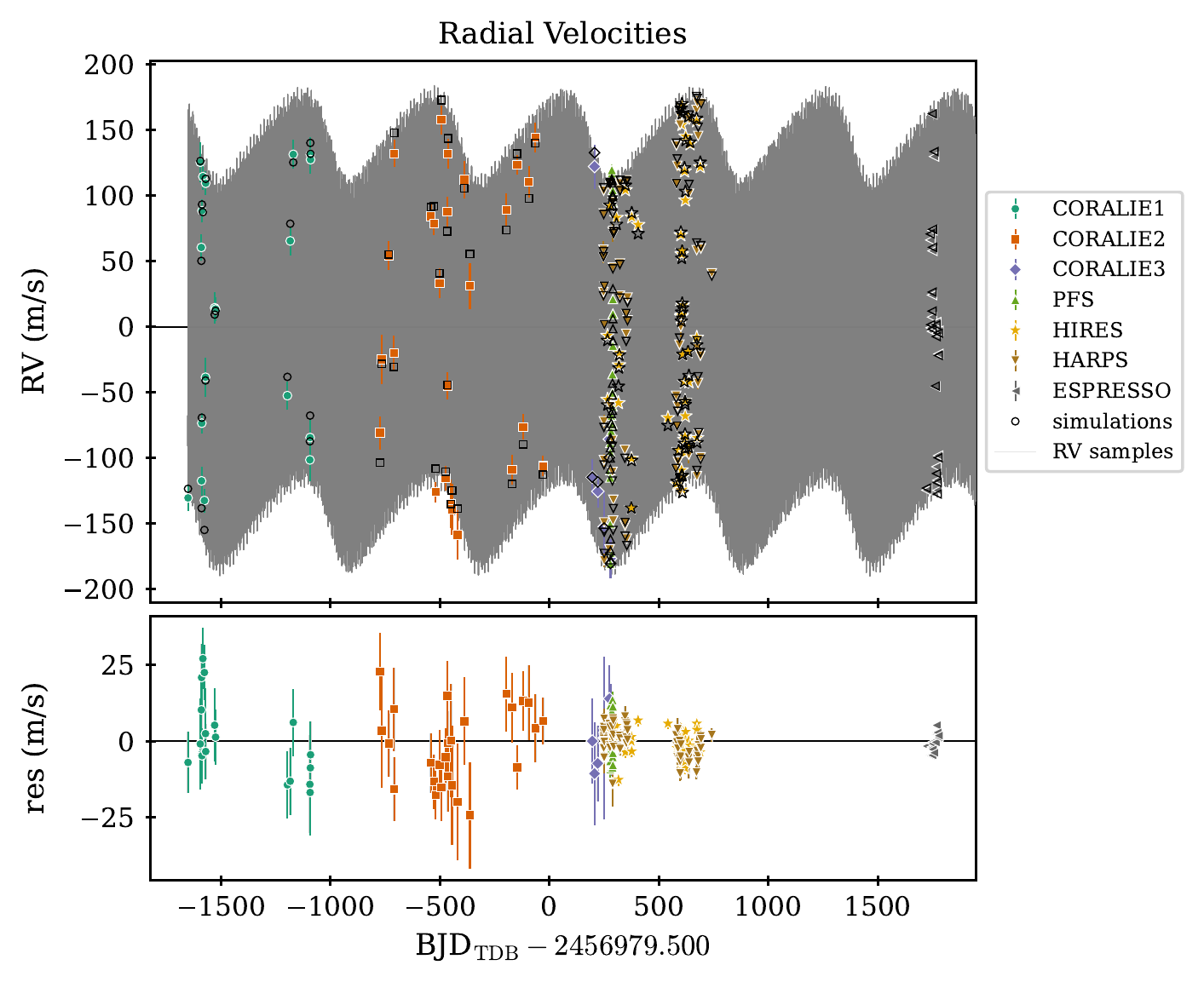}
    \caption{Radial velocities of WASP-47. Upper panel: RV plot. There is a different marker and color for each data set. The TRADES best-fit RV model is plotted with open black shapes. The maximum-likelihood orbital solution is shown by a light gray line. The CORALIE data set was divided into three data sets, as described in \citet{Almenara2016}. Lower panel: RV residuals with respect to the TRADES RV best-fit model. The corresponding jitter determined from the best-fit model has been added in quadrature to the measured uncertainty of each \valerio{data point}.}
    \label{fig:trades_rv}
\end{figure*}

In order to save computational time, we first ran a local minimizer\footnote{Nelder-Mead method implemented in the \texttt{scipy.optimize.min\-i\-mize} function.} and used the result as a starting point for the python package \texttt{emcee} \citep{Foreman2013, DFM2019JOSS....4.1864F}, for which we initialized 68 walkers with a tight Gaussian. We adopted as a sampler algorithm a mix of the differential evolution proposal \citep[80\% of the walkers;][]{DEMOVE2014ApJS..210...11N} and
the snooker differential evolution proposal \citep[20\% of the walkers;][]{terBraak2008}. We let the code run for 110\,000 steps 
and discarded the first 100\,000 steps as burn-in after checking the convergence of the chains by means of visual inspection as well as through diagnostics from \citealt{geweke1991} (within $2$-$3\sigma$) and Gelman-Rubin  statistics\footnote{The \^{R} statistic reached 1.01, but we note that this statistic is not effective with the sampling algorithm we used.} \citep{GelmanRubin1992}.
Our uncertainties are computed as the high density interval (HDI) at $68.27\%$ from the posterior distribution as the equivalent of the credible intervals at the $16$th and $84$th percentile. We defined our best-fit solution as the maximum likelihood estimator (MLE), that is, the parameter sample that maximizes the log-likelihood ($\log\mathcal{L}$) of the posterior distribution within the HDI. We also computed a symmetric uncertainty ($\sigma$) as the $68.27$th percentile of the (sorted) absolute residuals of the posterior with respect to the best-fit solution. (See Table~\ref{tab:trades_parameters} for a summary of the fitted and physical parameters of the system determined with TRADES. See the O-C diagrams of planet -b, -d, and -e in Figs.~\ref{fig:trades_oc_b}, \ref{fig:trades_oc_d},  and \ref{fig:trades_oc_e} and the RV plot in Fig~\ref{fig:trades_rv}.)

\begin{table*}
    \centering\renewcommand{\arraystretch}{1.3}
    \caption{Summary of the best-fit MLE parameters for the WASP-47 system determined with TRADES.}
    {
    \small
    \begin{tabular}{lcc}
    \hline\hline
    parameter (unit) & MLE (HDI \& $\sigma$)  & priors \\
    \hline
    \emph{planet b} & & \\
    $M_\mathrm{p}/M_\star\,(M_\odot/M_\star \times10^{-3})$ & $1.0617 _{-0.0007} ^{+0.0015} \pm 0.0012$ & \\
    $P$~(days)                                   & $ 4.158548 _{-0.000017} ^{+0.000008} \pm 0.000013 $  & $\mathcal{U}(3.8, 4.5)$\\        
    $\sqrt{e}\cos\omega$                         & $ 0.0005 _{-0.0091} ^{+0.0042} \pm 0.007 $          & \\             
    $\sqrt{e}\sin\omega$                         & $ 0.0239 _{-0.004} ^{+0.009} \pm 0.007 $          & \\
    $\lambda\,(^\circ)$                          & $ 329.05 _{-0.01} ^{+0.03} \pm 0.02 $                & $\mathcal{U}(0, 360)$ \\
    $M_\mathrm{p}\,(M_\oplus)$                   &$ 374 _{-16} ^{+17} \pm 17 $                           & $\mathcal{U}(80, 636)$ \\
    $e$                                          &$ 0.0006 _{-0.0002} ^{+0.0005} \pm 0.0004 $            & $\mathcal{U}(0.0, 0.5)$ \\
    $\omega\,(^\circ)$                           &$ 89 _{-11} ^{+19} \pm 17 $                            & $\mathcal{U}(0, 360)$ \\
    $\mathcal{M}\,(^\circ)$                      &$ 60 _{-19} ^{+11} \pm 17 $                            & $\mathcal{U}(0, 360)$ \\
    
    \emph{planet c} & & \rule{0pt}{15pt}\\
    $M_\mathrm{p}/M_\star\,(M_\odot/M_\star \times10^{-3})$ & $ 1.270 _{-0.017} ^{+0.005} \pm 0.012 $   & \\
    $P$~(days)                                   & $ 589.57 _{-0.02} ^{+0.02} \pm 0.02 $        & $\mathcal{U}(560, 600)$ \\
    $\sqrt{e}\cos\omega$                         & $ -0.19060 _{-0.02} ^{+0.00001} \pm 0.015 $     & \\
    $\sqrt{e}\sin\omega$                         & $  0.478 _{-0.016} ^{+0.007} \pm 0.014 $     & \\
    $\lambda\,(^\circ)$                          & $ 166.02 _{-0.01} ^{+0.03} \pm 0.02 $                & $\mathcal{U}(0, 360)$ \\
    $M_\mathrm{p}\,(M_\oplus)$                   & $ 447 _{-22} ^{+18} \pm 20 $                         & $\mathcal{U}(159, 636)$ \\
    $e$                                          & $ 0.264 _{-0.012} ^{+0.007} \pm 0.011 $              & $\mathcal{U}(0.0, 0.5)$ \\
    $\omega\,(^\circ)$                           & $ 111.8 _{-0.2} ^{+2.7} \pm 2 $                      & $\mathcal{U}(0, 360)$ \\
    $\mathcal{M}\,(^\circ)$                      & $ 234.3 _{-2.7} ^{+0.2} \pm 2 $                      & $\mathcal{U}(0, 360)$ \\
    
    \emph{planet d} & & \rule{0pt}{15pt}\\
    $M_\mathrm{p}/M_\star\,(M_\odot/M_\star \times10^{-3})$ & $ 0.0440 _{-0.0017} ^{+0.0007} \pm 0.0013 $ & \\
    $P$~(days)                                   & $ 9.09577 _{-0.00008} ^{+0.00016} \pm 0.00013 $            & $\mathcal{U}(8.7, 9.3)$ \\
    $\sqrt{e}\cos\omega$                         & $ 0.0037 _{-0.0115} ^{+0.0017} \pm 0.009 $            & \\
    $\sqrt{e}\sin\omega$                         & $ 0.031 _{-0.006} ^{+0.016} \pm 0.012 $            & \\
    $\lambda\,(^\circ)$                          & $ 278.61 _{-0.01} ^{+0.04} \pm 0.03 $            & $\mathcal{U}(0, 360)$ \\
    $i\,(^\circ)$                                & $ 90.805 _{-0.014} ^{+0.020} \pm 0.018 $            & $\mathcal{U}(80, 100)$ \\
    $\Omega\,(^\circ)$                           & $ 179.954 _{-0.012} ^{+0.006} \pm 0.010 $            & $\mathcal{U}(0, 360)$ \\
    $M_\mathrm{p}\,(M_\oplus)$                   & $ 15.5 _{-0.9} ^{+0.7} \pm 0.8 $                & $\mathcal{U}(1, 159)$ \\
    $e$                                          & $ 0.0010 _{-0.0007} ^{+0.0008} \pm 0.0008 $                & $\mathcal{U}(0.0, 0.5)$ \\
    $\omega\,(^\circ)$                           & $ 83 _{-3} ^{+18} \pm 14 $                & $\mathcal{U}(0, 360)$\\
    $\mathcal{M}\,(^\circ)$                      & $ 16 _{-15} ^{+10} \pm 14 $                & $\mathcal{U}(0, 360)$\\
    
    \emph{planet e} & \rule{0pt}{15pt}\\
    $M_\mathrm{p}/M_\star\,(M_\odot/M_\star \times10^{-3})$ & $ 0.0254 _{-0.0003} ^{+0.0010} \pm 0.0008 $ & \\
    $P$~(days)                                   & $ 0.789608 _{-0.000001} ^{+0.000002} \pm 0.000001 $    & $\mathcal{U}(0.7, 0.86)$ \\
    $\lambda\,(^\circ)$                          & $ 149.074 _{-0.033} ^{+0.015} \pm 0.03 $              & $\mathcal{U}(0, 360)$ \\
    $M_\mathrm{p}\,(M_\oplus)$                   & $ 9.0 _{-0.4} ^{+0.6} \pm 0.5 $               & $\mathcal{U}(1, 80)$ \\
    $\mathcal{M}\,(^\circ)$                      & $ 244.15 _{-0.033} ^{+0.015} \pm 0.025 $               & $\mathcal{U}(0, 360)$ \\
    
    RV data-sets & & \rule{0pt}{15pt}\\
    
    $\gamma_\mathrm{CORALIE1}\,(\mathrm{ms^{-1}})$ & $ -27069 _{-2} ^{+3} \pm 2 $ & \\
    $\gamma_\mathrm{CORALIE2}\,(\mathrm{ms^{-1}})$ & $ -27084.9 _{-4.2} ^{+0.4} \pm 3.2 $ & \\
    $\gamma_\mathrm{CORALIE3}\,(\mathrm{ms^{-1}})$ & $ -27065 _{-2} ^{+8} \pm 6 $ & \\
    $\gamma_\mathrm{PFS}\,(\mathrm{ms^{-1}})$      & $ 17.80 _{-0.09} ^{+1.20} \pm 0.8 $ & \\
    $\gamma_\mathrm{HIRES}\,(\mathrm{ms^{-1}})$    & $ 4.6 _{-0.6} ^{+0.2} \pm 0.5 $ & \\
    $\gamma_\mathrm{HARPS}\,(\mathrm{ms^{-1}})$    & $ -27041.1 _{-0.3} ^{+0.4} \pm 0.3 $ & \\
    $\gamma_\mathrm{ESPRESSO}\,(\mathrm{ms^{-1}})$ & $ -27165.7 _{-0.5} ^{+0.1} \pm 0.3 $ & \\
\end{tabular}
    }
    \tablefoot{The columns give the name of the parameter, its best-fit value with its associated $68.27\%$ HDI and $\sigma$, and the priors adopted.
    Astrocentric orbital parameters are defined at $2456979.5\, \mathrm{BJD_{TDB}}$.
    $\mathcal{M}$ is the mean anomaly computed as $\mathcal{M} = \lambda - \omega - \Omega$.
    We fixed the value of $i_\mathrm{c}=90^\circ$, so $M_\mathrm{p}$ of planet -c is the minimum mass.
    Planet -e has a circular orbit ($e=0, \omega=90^\circ)$ at the reference time, so $e$ amd $\omega$ have not been fitted with TRADES. See Section \ref{sec:analysis_dynamical} for details.}
    \label{tab:trades_parameters}
\end{table*}

\section{Discussion and conclusions}\label{sec:discussion_conclusions}

The global analysis presented in this work is the most comprehensive dynamical modeling of the WASP-47 planetary system carried out so far. We exploited all the available RV and TTV data sets and merged them with 19 unpublished light curves, 11 of which were observed with CHEOPS. Our newly derived planetary masses ($M_b$, $M_c$, $M_d$, and $M_e$) from the final best-fit parameters (Table~\ref{tab:trades_parameters}) can be compared with the values reported in the literature and listed in Table~\ref{tab:masses}. 

\begin{figure}
    \centering
    \includegraphics[width=\columnwidth]{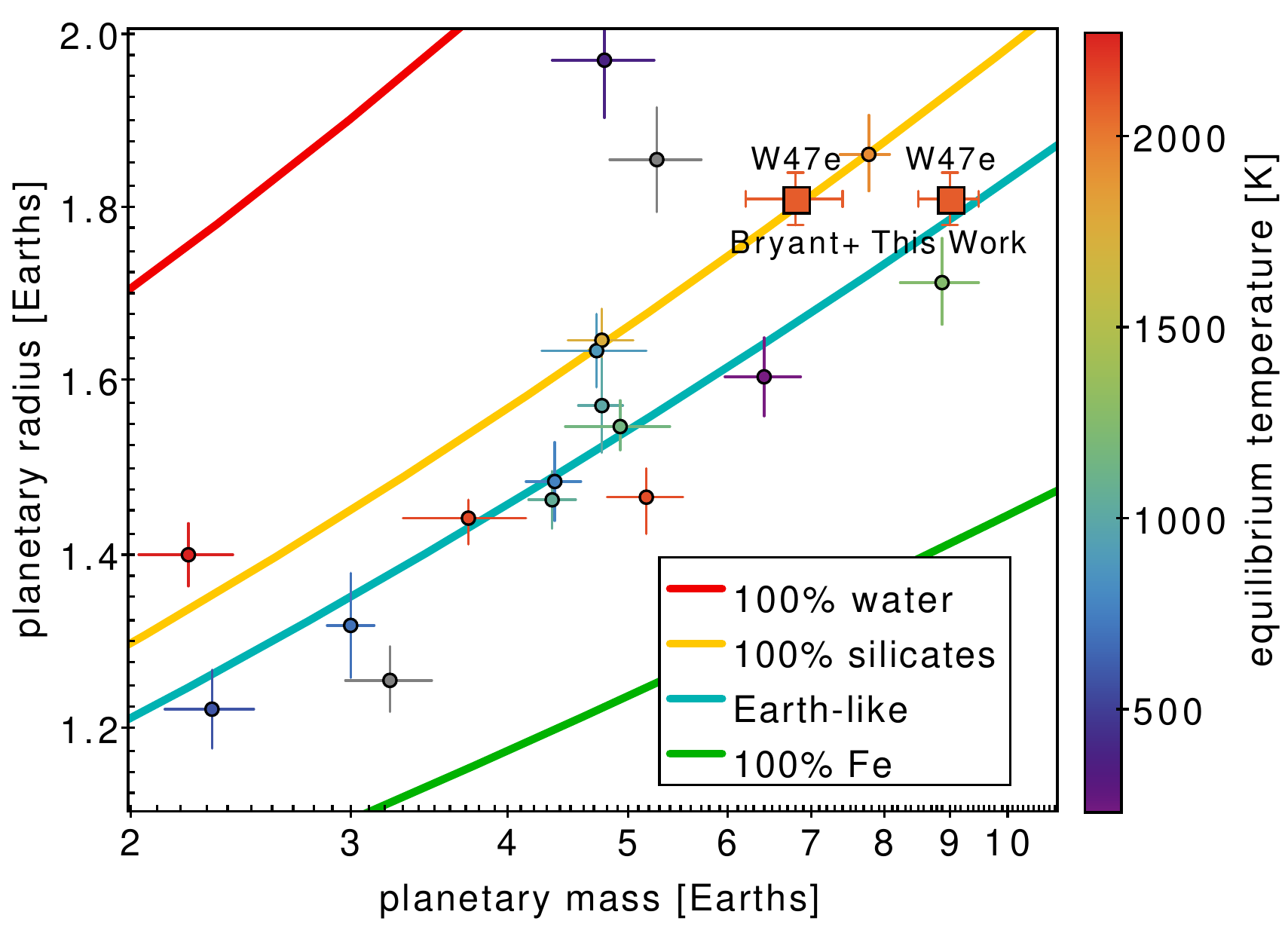}
    \caption{Mass-radius diagram with all the known planets with a relative error on bulk density better than 20\% (circle points) color coded according to their equilibrium temperature. The colored lines represent models at different bulk compositions from \citet{Zeng2019}. The two measurements of WASP-47e by us and by \citet{Bryant2022} are plotted as squares.}
    \label{fig:mr}
\end{figure}

For the  two low-mass planets (-d, -e), we found best-fit masses ($M_d/M_\oplus=15.5\pm0.8$,  $M_e/M_\oplus=9.0\pm0.5$) with relative errors of about 5\%, which is smaller than all the previous relative errors and is now limited mostly by our knowledge of the stellar mass (relative error on $M_\star$: $\sim $4\%). In particular, for the innermost planet, our solution is perfectly consistent with the upper extreme $M_e/M_\oplus=9.1\pm1.0$ set by \citealt{Weiss2017} (joint RV plus TTV analysis), but it is not consistent with the lower extreme $M_e/M_\oplus=6.8\pm0.6$ set by \citealt{Bryant2022} using RVs only, apparently confirming, at 2.8$\sigma$ significance, some kind of systematic offset between the two techniques \citep{Mills2017,Petigura2018}. We did not see a similar behavior for planet -d, whose best-fit mass is fully consistent with all the previous measurements but more precise. By calculating the bulk densities as derived quantities from Table~\ref{tab:trades_parameters} and propagating the errors, we get $\rho_\mathrm{d} = 1.69 \pm 0.22\, \mathrm{g\,cm^{-3}}$ and $\rho_\mathrm{e} = 8.1 \pm 0.5\, \mathrm{g\,cm^{-3}}$, thus confirming a Neptune-like density and composition for planet -d (implying an extended volatile envelope) but also moving planet -e very close to what planetary structure models predict to be an Earth-like composition rather than a pure-silicate one (in contrast to the claims by \citealt{Vanderburg2017} and \citealt{Bryant2022}; see Fig.~\ref{fig:mr}). If that conclusion were true, then planet -e would not appear as an outlier in the planetary density versus~stellar irradiation diagram anymore and there would be no reason to hypothesize a different formation path with respect to other ultra-short period planets at similar irradiation levels, such as K2-141b and HD~213885b \citep{Malavolta2018,Espinoza2020}. Moreover, WASP-47 is a rather metal-rich star at $\textrm{[Fe/H]} = +0.36\pm 0.05$, according to \citet{Mortier2013}. The high density of WASP-47e, which fits into the ``super-Mercury'' class, would confirm the finding by \citet{Adibekyan2021} that the bulk density of super-Earths correlates with stellar iron fraction. On the other hand, the relatively low density of planet -d would also confirm the trend discovered by \citet{Wilson2022}, who found that sub-Neptunes around metal-rich stars have, on average, lower densities. Taken together, our findings would support the general idea that a higher stellar metallicity leads to forming planets with larger cores and, hence, generates denser hot super-Earths (such as -e) and warm sub-Neptunes that are able to retain more extended atmospheric envelopes (-d).

We caution, however, that the existing tension on $M_\mathrm{e}$ could be due to subtle methodological biases in RV and/or RV plus TTV analyses that are yet to be fully explored and a deeper understanding of the effects at play in such dynamically complex systems is needed. This is true of course for not only WASP-47 but for all planetary systems for which both techniques can be applied and that are steadily growing in number, especially after the launch of K2 and TESS. Two factors are most frequently discussed when dealing with such issues: the impact of stellar activity and the way RV and TTV information is merged at the analysis stage. Notably, WASP-47 is not a particularly active star, and all the studies up to and including \citet{Vanderburg2017} explicitly neglected any contribution from stellar activity on both photometric and spectroscopic data. \citet{Bryant2022}, on the other hand, noticed an excess scatter in the ESPRESSO data and concluded that the inclusion of a Gaussian process kernel in their RV modeling was justified by a lower BIC value of their fit, even though the statistical significance of the peak on the periodogram of the RV residuals at the claimed rotational period is marginal. The treatment of stellar activity, however, does not appear to be the (only) explanation for the $M_\mathrm{e}$ discrepancy we see since \citet{Bryant2022} came to a value that is almost identical to that reported by \citet{Vanderburg2017} without modeling for it: $M_\mathrm{e} = 6.8\pm0.7$ vs.~$6.8\pm0.6$~$M_\oplus$. Beyond that, in an RV plus TTV study such ours, it is extremely difficult to model stellar activity in a consistent framework because the two techniques are effected in different ways and at different timescales \citep{Boisse2011,Oshagh2013,Ioannidis2016}; indeed, transit light curves are impacted more by the local rather than global distribution of star spots over the photosphere. With WASP-47, an additional issue prevents us from adopting a more complicated data model (i.e., including stellar activity and/or a photodynamical approach): computational time. Having to deal with four planets, 34 free parameters, and an extensive set of 133 transits and 212 RVs, each of the MCMC steps of the optimization process described in Section~\ref{sec:analysis_dynamical} took about 15~s on a medium-power computing workstation, and the whole fit required weeks to reach convergence. This sizable computing time is mostly due to an unfortunate combination of a very long observing baseline (15~years) and the very short period of the inner planet ($P_\mathrm{e}\simeq 0.79$~d), forcing the N-body integration time step to unusually small values. Of course, under these assumptions, the inclusion of a Gaussian process treatment of stellar activity or, even worse, the implementation of a photodynamical algorithm would increase the computational time up to an unreasonable amount, at least with ordinarily available hardware. We emphasize, however, that our result agrees with previous RV plus TTV studies both based \citep{Almenara2016} and not based \citep{Weiss2017} on a photodynamical approach, so fitting our light curves and performing the dynamical modeling at separate stages cannot be the sole reason for the discrepancy discussed above.

As for the giant planets (-b and -c), it is worth noting that our newly derived masses are compatible with those presented by \citet{Vanderburg2017} and \citet{Bryant2022} and  statistically consistent with all the previous measurements reported in Table~\ref{tab:masses}. The precision of our values $M_b/M_\oplus=374\pm17$, $M_c/M_\oplus=447\pm 20$ is larger by a factor of about two with respect to the most recent RV works. This is easily understood if we consider that our error on $M_\star$ ($4.4\%$) is larger than theirs (3\%) and our dynamical analysis probes are $M_\mathrm{p}/M_\star$ rather than $M_\mathrm{p}/M_\star^{2/3}$, the latter being true for a pure RV analysis. In other words, our study is more sensitive to the uncertainty on stellar mass. The same is true for the derived planetary density $\rho_\mathrm{b} = 0.98 \pm 0.09$~$\mathrm{g\, cm^{-3}}$.

An important byproduct of our analysis is the derivation of new, updated mean ephemerides\footnote{It is worth noting that the mean ephemeris, by definition, does not account for the TTV modulation. The latter will manifest itself as a (quasi-)periodic term added to the ephemeris with amplitudes $A_\mathrm{TTV,b}\simeq 0.7$~min and $A_\mathrm{TTV,d}\simeq 5.8$~min.} for all the WASP-47 transiting planets, as this data can be exploited by any future follow-up study. Our best-fit relations are:
\begin{eqnarray}
T_\mathrm{0,b} &=& 2459407.761987 \pm 0.0000002 \textrm{ BJD}_\textrm{TDB}\\
& & + N\times (4.159151 \pm 0.0000004),\nonumber
\end{eqnarray}
\begin{eqnarray}
T_\mathrm{0,d} &=& 2457024.430107 \pm 0.0000003 \textrm{ BJD}_\textrm{TDB}\\
& & + N\times (9.030501 \pm 0.000013),\nonumber
\end{eqnarray}
\begin{eqnarray}
T_\mathrm{0,e} &=& 2457012.13666 \pm 0.00003 \textrm{ BJD}_\textrm{TDB}\\
& & + N\times (0.78961 \pm 0.00003),\nonumber
\end{eqnarray}
where $N$ is an integer number commonly known as ``epoch'' and set arbitrarily to zero at our reference transit time $T_\mathrm{ref}$. We highlight that the 1$\sigma$ uncertainty on transit prediction at 2023.0 for planet -d has now been reduced from about one hour to just six minutes. Of course this is crucial for planning space-based observations, where the cost of the invested observing time and the time-sensitive nature of the observations makes pinpointing the transit window as precisely as possible critical. 

Observations of WASP-47 are not yet scheduled with JWST, at least during Cycle 1, even though its transiting planets -e, -b, and -d have been recognized as suitable targets for transmission spectroscopy \citep{Bryant2022} and will very likely be proposed in the next cycles. Such observations will be crucial to test our planetary formation theories. It is widely accepted that three main channels for HJ migration exist \citep{Dawson2018}: one is through the protoplanetary disk at low eccentricity, another is through high-eccentricity dynamical migration with tidal damping at later times, and the third is through in situ formation close to or at the final orbit. A planet migrating through the disc or having formed in situ would accrete a large amount of gas from inside the ice line, whereas an HJ that migrated dynamically may have originated from beyond the ice line with a different composition \citep{Oberg2011,Madhu2017,Knierim2022}. Simulations show that the dynamical migration channel would invariably destroy any close-in planets, as the orbit of the HJ would intersect with that of the close-in planets prior to circularization \citep{Mustill2015}. Thus, unlike ordinary HJs, for very rare systems such as WASP-47, we can rule out one of the main scenarios and test the other models through detailed atmospheric characterization. 

Other space-based follow-up opportunities include, at least in principle, TESS, PLATO \citep{Rauer2014}, and ARIEL \citep{Tinetti2018}. As for TESS, we mentioned in the introduction that no other sector will include WASP-47, at least up to the end of Cycle 6, implying that this target will not be re-observed until fall 2024, at best. Unfortunately, individual transits of planets -e and -d are essentially undetectable by TESS due to its noise properties, and the time baseline of each sector is limited to about four weeks. Future sectors, if any, will thus not be  effective in significantly extending the TTV analysis carried out so far.

On the other hand, PLATO will be able to deliver high-precision transit light curves of planets -e and -d since the predicted noise level ranges from 90 to 180~ppm in one hour, according to the number of co-pointing PLATO cameras involved \citep{Montalto2021}. Due to its very low ecliptic latitude, WASP-47 will not lie within the proposed Long-duration Observing Phase fields of PLATO \citep{Nascimbeni2022}, where targets will be continuously observed for at least two years. However, WASP-47 could be selected for the Short-duration Observing Phase and monitored for two to three months without any significant interruption (see Fig.~1 of \citealt{Heller2022}). 

As a closing note, ARIEL, though focused on the atmospheric characterization of exoplanets through emission and transmission spectroscopy \citep{Tinetti2021}, has also been proposed as a very powerful instrument to investigate TTVs \citep{Borsato2021a}. WASP-47 is consistently included in the provisional ARIEL target list \citep{Zingales2018,Edwards2022} as a ``Tier 2'' object (aim: in-depth atmospheric characterization through repeated observations), but these observations are only devoted to observing eclipses of planet -b. 

\begin{acknowledgements}
CHEOPS is an ESA mission in partnership with Switzerland with important contributions to the payload and the ground segment from Austria, Belgium, France, Germany, Hungary, Italy, Portugal, Spain, Sweden, and the United Kingdom. The CHEOPS Consortium would like to gratefully acknowledge the support received by all the agencies, offices, universities, and industries involved. Their flexibility and willingness to explore new approaches were essential to the success of this mission. 
LBo, GBr, VNa, IPa, GPi, RRa, GSc, VSi, and TZi acknowledge support from CHEOPS ASI-INAF agreement n. 2019-29-HH.0. 
FZM is funded by ``Bando per il Finanziamento di Assegni di Ricerca Progetto Dipartimenti di Eccellenza Anno 2020'' and is co-funded in agreement with ASI-INAF n.2019-29-HH.0 from 26 Nov/2019 for ``Italian participation in the operative phase of CHEOPS mission'' (DOR - Prof.~Piotto).
This project has received funding from the European Research Council (ERC) under the European Union’s Horizon 2020 research and innovation programme (project {\sc Four Aces}. 
SH gratefully acknowledges CNES funding through the grant 837319. 
S.G.S. acknowledge support from FCT through FCT contract nr. CEECIND/00826/2018 and POPH/FSE (EC). 
ACC and TW acknowledge support from STFC consolidated grant numbers ST/R000824/1 and ST/V000861/1, and UKSA grant number ST/R003203/1. 
V.V.G. is an F.R.S-FNRS Research Associate. 
The Belgian participation to CHEOPS has been supported by the Belgian Federal Science Policy Office (BELSPO) in the framework of the PRODEX Program, and by the University of Liège through an ARC grant for Concerted Research Actions financed by the Wallonia-Brussels Federation. 
L.D. is an F.R.S.-FNRS Postdoctoral Researcher. 
YA and MJH acknowledge the support of the Swiss National Fund under grant 200020\_172746. 
We acknowledge support from the Spanish Ministry of Science and Innovation and the European Regional Development Fund through grants ESP2016-80435-C2-1-R, ESP2016-80435-C2-2-R, PGC2018-098153-B-C33, PGC2018-098153-B-C31, ESP2017-87676-C5-1-R, MDM-2017-0737 Unidad de Excelencia Maria de Maeztu-Centro de Astrobiologí­a (INTA-CSIC), as well as the support of the Generalitat de Catalunya/CERCA programme. The MOC activities have been supported by the ESA contract No. 4000124370. 
S.C.C.B. acknowledges support from FCT through FCT contracts nr. IF/01312/2014/CP1215/CT0004. 
XB, SC, DG, MF and JL acknowledge their role as ESA-appointed CHEOPS science team members. 
ABr was supported by the SNSA. 
ACC acknowledges support from STFC consolidated grant numbers ST/R000824/1 and ST/V000861/1, and UKSA grant number ST/R003203/1. 
This project was supported by the CNES. 
This work was supported by FCT - Fundação para a Ciência e a Tecnologia through national funds and by FEDER through COMPETE2020 - Programa Operacional Competitividade e Internacionalizacão by these grants: UID/FIS/04434/2019, UIDB/04434/2020, UIDP/04434/2020, PTDC/FIS-AST/32113/2017 \& POCI-01-0145-FEDER- 032113, PTDC/FIS-AST/28953/2017 \& POCI-01-0145-FEDER-028953, PTDC/FIS-AST/28987/2017 \& POCI-01-0145-FEDER-028987, O.D.S.D. is supported in the form of work contract (DL 57/2016/CP1364/CT0004) funded by national funds through FCT. 
B.-O. D. acknowledges support from the Swiss State Secretariat for Education, Research and Innovation (SERI) under contract number MB22.00046. 
MF and CMP gratefully acknowledge the support of the Swedish National Space Agency (DNR 65/19, 174/18).
AJM acknowledges support from the Swedish National Space Agency (career grant 120/19C).
DG gratefully acknowledges financial support from the CRT foundation under Grant No. 2018.2323 ``Gaseousor rocky? Unveiling the nature of small worlds''. 
M.G. is an F.R.S.-FNRS Senior Research Associate. 
KGI is the ESA CHEOPS Project Scientist and is responsible for the ESA CHEOPS Guest Observers Programme. She does not participate in, or contribute to, the definition of the Guaranteed Time Programme of the CHEOPS mission through which observations described in this paper have been taken, nor to any aspect of target selection for the programme. 
This work was granted access to the HPC resources of MesoPSL financed by the Region Ile de France and the project Equip@Meso (reference ANR-10-EQPX-29-01) of the programme Investissements d'Avenir supervised by the Agence Nationale pour la Recherche. 
ML acknowledges support of the Swiss National Science Foundation under grant number PCEFP2\_194576. 
PM acknowledges support from STFC research grant number ST/M001040/1. 
This work was also partially supported by a grant from the Simons Foundation (PI Queloz, grant number 327127). 
IRI acknowledges support from the Spanish Ministry of Science and Innovation and the European Regional Development Fund through grant PGC2018-098153-B- C33, as well as the support of the Generalitat de Catalunya/CERCA programme. 
GyMSz acknowledges the support of the Hungarian National Research, Development and Innovation Office (NKFIH) grant K-125015, a a PRODEX Experiment Agreement No. 4000137122, the Lend\"ulet LP2018-7/2021 grant of the Hungarian Academy of Science and the support of the city of Szombathely. 
NAW acknowledges UKSA grant ST/R004838/1. \\
This work was supported by FCT - Fundação para a Ciência e a Tecnologia through national funds and by FEDER through COMPETE2020 - Programa Operacional Competitividade e Internacionalização by these grants: UID/FIS/04434/2019; UIDB/04434/2020; UIDP/04434/2020. 
DB has been partially funded by MCIN/AEI/10.13039/501100011033 grants PID2019-107061GB-C61 and MDM-2017-0737.
%
This project has received funding from the European Research Council (ERC) under the European Union’s Horizon 2020 research and innovation programme (project {\sc Four Aces}; grant agreement No 724427). It has also been carried out in the frame of the National Centre for Competence in Research PlanetS supported by the Swiss National Science Foundation (SNSF). DE acknowledges financial support from the Swiss National Science Foundation for project 200021\_200726.
P.E.C. is funded by the Austrian Science Fund (FWF) Erwin Schroedinger Fellowship, program J4595-N.
S.~S. have received funding from the European Research Council (ERC) under the European Union’s Horizon 2020 research and innovation program (grant agreement No 833925, project STAREX).

This research has made use of the SIMBAD database (operated at CDS, Strasbourg, France; \citealt{Wenger2000}), the VARTOOLS Light Curve Analysis Program (version 1.39 released October 30, 2020, \citealt{Hartman_and_Bakos_2016}), TOPCAT and STILTS \citep{Taylor2005,Taylor2006}, the NASA Exoplanet archive \citep{Akeson2013}. Based on observations made with the REM Telescope, INAF Chile.

\end{acknowledgements}

\bibliographystyle{aa}
\bibliography{biblio}

\begin{appendix}
\label{sec:appendix}

\clearpage

\onecolumn

\section{Data tables}
Transits of planets WASP-47b, WASP-47d, WASP-47e fitted in the present work (Section~\ref{sec:analysis_lc}) and included in our global dynamical analysis (Section~\ref{sec:analysis_dynamical}).

\begin{table*}[!h]
    \centering\renewcommand{\arraystretch}{1.3}
    \caption{Transits of planet WASP-47b from space-based telescopes included in our global analysis.}
    \begin{tabular}{cccccccc} \hline\hline
      source & UT date & N & cadence & $\sigma_\mathrm{phot}$ & $T_0$ & $\sigma(T_0)$ & $\sigma(T_0)$ \\
      (+ band) & (of $T_0$) & & (s)  & (ppm) & ($\mathrm{BJD_{TDB}}$) & (days) & (s) \\ \hline
  CHEOPS & 2020-06-24 & 242 & 60 & 1050  & 2459025.11931 & 0.00054 & 47 \rule{0pt}{15pt}\\
  CHEOPS & 2020-07-07 & 251 & 60 & 1960  & 2459037.59764 & 0.00048 & 42 \\
  CHEOPS & 2020-10-02 & 283 & 60 & 1190  & 2459124.93989 & 0.00031 & 26 \\
  CHEOPS & 2021-07-20 & 271 & 60 & 1130  & 2459416.08220 & 0.00083 & 72 \\
  CHEOPS & 2021-07-28 & 236 & 60 & 1140  & 2459424.39829 & 0.00053 & 46 \\
  CHEOPS & 2021-08-18 & 522 & 60 & 1270  & 2459445.19505 & 0.00033 & 29 \\
  CHEOPS & 2021-08-27 & 537 & 60 & 1140  & 2459453.51274 & 0.00038 & 33 \\
  CHEOPS & 2021-09-12 & 357 & 60 & 1220  & 2459470.14844 & 0.00035 & 30 \\
  CHEOPS & 2021-09-16 & 512 & 60 & 1390  & 2459474.30828 & 0.00034 & 29 \\
  CHEOPS & 2021-09-25 & 247 & 60 & 2630  & 2459482.62705 & 0.00085 & 73 \\ 
    TESS & 2021-08-22 & 412 & 120 & 2150 & 2459449.35394 & 0.00079 & 68 \rule{0pt}{15pt}\\ 
    TESS & 2021-08-27 & 411 & 120 & 3900 & 2459453.51314 & 0.00132 & 114 \\
    TESS & 2021-08-31 & 319 & 120 & 3330 & 2459457.66953 & 0.00107 & 93 \\
    TESS & 2021-09-04 & 412 & 120 & 2040 & 2459461.83102 & 0.00080 & 69 \\
    TESS & 2021-09-08 & 411 & 120 & 2040 & 2459465.98830 & 0.00076 & 66 \\
    TESS & 2021-09-12 & 412 & 120 & 2110 & 2459470.14753 & 0.00083 & 72 \\ 
      K2 & 2014-11-21 & 741 & 60 & 347   & 2456982.97692 & 0.00043 & 37 \rule{0pt}{15pt}\\
      K2 & 2014-11-25 & 754 & 60  & 342  & 2456987.13597 & 0.00008 & 7  \\
      K2 & 2014-11-29 & 821 & 60 & 357   & 2456991.29538 & 0.00008 & 7  \\
      K2 & 2014-12-03 & 732 & 60 & 342   & 2456995.45461 & 0.00022 & 19 \\
      K2 & 2014-12-08 & 740 & 60 & 349   & 2456999.61388 & 0.00009 & 8  \\
      K2 & 2014-12-12 & 833 & 60 & 348   & 2457003.77342 & 0.00008 & 7  \\
      K2 & 2014-12-16 & 739 & 60 & 352   & 2457007.93265 & 0.00008 & 7  \\
      K2 & 2014-12-20 & 709 & 60 & 354   & 2457012.09155 & 0.00043 & 37 \\
      K2 & 2014-12-24 & 709 & 60 & 362   & 2457016.25091 & 0.00008 & 7  \\
      K2 & 2014-12-28 & 828 & 60 & 354   & 2457020.40989 & 0.00008 & 7  \\
      K2 & 2015-01-02 & 449 & 60 & 361   & 2457024.56863 & 0.00037 & 32 \\
      K2 & 2015-01-06 & 728 & 60 & 361   & 2457028.72772 & 0.00008 & 7  \\
      K2 & 2015-01-10 & 729 & 60 & 362   & 2457032.88682 & 0.00008 & 7  \\
      K2 & 2015-01-14 & 860 & 60 & 346   & 2457037.04569 & 0.00008 & 7  \\
      K2 & 2015-01-18 & 741 & 60 & 352   & 2457041.20502 & 0.00008 & 7  \\
      K2 & 2015-01-22 & 739 & 60 & 345   & 2457045.36340 & 0.00041 & 36 \\ \hline
    \end{tabular}
    \tablefoot{The columns give the instrument name, the UT date closest to the transit center $T_0$, the number of photometric data points, the average cadence of the light curve in seconds, the scatter of the residuals from our best-fit model in parts per million, the transit time $T_0$ in the $\textrm{BJD}_\textrm{TDB}$ standard and its associated error, and the latter converted into seconds. The fitting procedure is described in detail in Section~\ref{sec:analysis_lc}.}
    \label{tab:tzero_b_space}
\end{table*}

\FloatBarrier

\begin{table*}
    \centering\renewcommand{\arraystretch}{1.3}
    \caption{Transits of planet WASP-47b from ground-based telescopes included in our global analysis.}
    \begin{tabular}{lccccccc} \hline\hline
          source & UT date & N & cadence & $\sigma_\mathrm{phot}$ & $T_0$ & $\sigma(T_0)$ & $\sigma(T_0)$ \\
          (+ band) & (of $T_0$) & & (s)  & (ppm) & ($\mathrm{BJD_{TDB}}$) & (days) & (s) \\ \hline
        REM ($g'$)      &       2021-09-25      &       137     &       60      &       3002    &       2459482.61821   &       0.00197 &       171\rule{0pt}{15pt}         \\ 
        REM ($r'$)      &       2021-09-25      &       139     &       60      &       2119    &       2459482.62461   &       0.00106 &       91         \\ 
        REM ($i'$)      &       2021-09-25      &       139     &       60      &       2672    &       2459482.62212   &       0.00145 &       125         \\ 
        ETD     ($R$) & 2020-08-30      &       145     &       60      &       1315    &       2459091.66670   &       0.00064 &       56\rule{0pt}{15pt}         \\ 
        ETD     ($R$) & 2021-07-12      &       144     &       60      &       1581    &       2459407.76213   &       0.00082 &       71         \\ 
        ETD     ($R$) & 2021-08-06      &       141     &       60      &       1658    &       2459432.71923   &       0.00081 &       70         \\ 
        ETD     ($R$) & 2021-08-31      &       155     &       60      &       1617    &       2459457.67291   &       0.00077 &       67         \\ 
        ETD     ($R$) & 2021-09-25      &       147     &       60      &       1540    &       2459482.62786   &       0.00073 &       63         \\ 
        ETD (clear)     &       2021-09-25      &       142     &       60      &       1723    &       2459482.62868   &       0.00084 &       72         \\ 
        ETD     ($R$) & 2021-10-20      &       129     &       60      &       1730    &       2459507.58050   &       0.00085 &       74         \\ \hline
    \end{tabular}
    \tablefoot{The columns give the instrument name and pass band, the UT date closest to the transit center $T_0$, the number of photometric data points, the average cadence of the light curve in seconds, the scatter of the residuals from our best-fit model in parts per million, the transit time $T_0$ in the $\textrm{BJD}_\textrm{TDB}$ standard and its associated error, and the latter converted into seconds. The fitting procedure is described in detail in Section~\ref{sec:analysis_lc}.}
    \label{tab:tzero_b_ground}
\end{table*}

\begin{table*}
    \centering\renewcommand{\arraystretch}{1.3}
    \caption{Transits of planet WASP-47d included in our global analysis.}
    \begin{tabular}{cccccccc} \hline\hline
        source & UT date & N & cadence & $\sigma_\mathrm{phot}$ & $T_0$ & $\sigma(T_0)$ & $\sigma(T_0)$ \\
        (+ band) & (of $T_0$) & & (s)  & (ppm) & ($\mathrm{BJD_{TDB}}$) & (days) & (s) \\ \hline
        CHEOPS & 2020-06-24 & 242 & 60 & 1050 & 2459025.11931 & 0.00054 & 47\rule{0pt}{15pt}\\
        K2 & 2014-11-21 & 741 & 60 & 347 & 2456982.97692 & 0.00043 & 37\rule{0pt}{15pt}\\
        K2 & 2014-11-25 & 754 & 60  & 342 & 2456987.13597 & 0.00008 & 7\\
        K2 & 2014-11-29 & 821 & 60 & 357 & 2456991.29538 & 0.00008 & 7\\
        K2 & 2014-12-03 & 732 & 60 & 342 & 2456995.45461 & 0.00022 & 19\\
        K2 & 2014-12-08 & 740 & 60 & 349 & 2456999.61388 & 0.00009 & 8\\
        K2 & 2014-12-12 & 833 & 60 & 348 & 2457003.77342 & 0.00008 & 7\\
        K2 & 2014-12-16 & 739 & 60 & 352 & 2457007.93265 & 0.00008 & 7\\
        K2 & 2014-12-20 & 709 & 60 & 354 & 2457012.09155 & 0.00043 & 37\\
        K2 & 2014-12-24 & 709 & 60 & 362 & 2457016.25091 & 0.00008 & 7\\
        K2 & 2014-12-28 & 828 & 60 & 354 & 2457020.40989 & 0.00008 & 7\\
        K2 & 2015-01-02 & 449 & 60 & 361 & 2457024.56863 & 0.00037 & 32\\
    \end{tabular}
    \tablefoot{The columns give the instrument name, the UT date closest to the transit center $T_0$, the number of photometric data points, the average cadence of the light curve in seconds, the scatter of the residuals from our best-fit model in parts per million, the transit time $T_0$ in the $\textrm{BJD}_\textrm{TDB}$ standard and its associated error, and the latter converted into seconds. The fitting procedure is described in detail in Section~\ref{sec:analysis_lc}.}
    \label{tab:tzero_d}
\end{table*}

\begin{table*}
    \centering\renewcommand{\arraystretch}{1.3}
    \caption{CHEOPS data sets used in our analysis.}
    \begin{tabular}{c} \hline\hline
    \texttt{CH\_PR100025\_TG005601\_V0200}\hspace{5mm}
\texttt{CH\_PR100025\_TG005602\_V0200}\hspace{5mm}
\texttt{CH\_PR100017\_TG000101\_V0200}\\
\texttt{CH\_PR100025\_TG005603\_V0200}\hspace{5mm}
\texttt{CH\_PR100025\_TG006201\_V0200}\hspace{5mm}
\texttt{CH\_PR100025\_TG003601\_V0200}\\
\texttt{CH\_PR100025\_TG006301\_V0200}\hspace{5mm}
\texttt{CH\_PR100025\_TG006302\_V0200}\hspace{5mm}
\texttt{CH\_PR100025\_TG003602\_V0200}\\
\texttt{CH\_PR100025\_TG006303\_V0200}\hspace{5mm}
\texttt{CH\_PR100025\_TG003603\_V0200}\hspace{5mm}
\phantom{\texttt{CH\_PR100025\_TG003603\_V0200}}\\ \hline
    \end{tabular}
    \label{tab:datasets}
\end{table*}

\begin{table*}
    \centering\renewcommand{\arraystretch}{1.25}
    \caption{Transits of planet WASP-47e included in our global analysis, part one.}
    \begin{tabular}{cccccccc} \hline\hline
        source & UT date & N & cadence & $\sigma_\mathrm{phot}$ & $T_0$ & $\sigma(T_0)$ & $\sigma(T_0)$ \\
        (+ band) & (of $T_0$) & & (s)  & (ppm) & ($\mathrm{BJD_{TDB}}$) & (days) & (s) \\ \hline
        K2      &       2014-11-18      &       205     &       60      &       254     &       2456979.75947   &       0.00304 &       262         \\ 
        K2      &       2014-11-19      &       200     &       60      &       264     &       2456980.54893   &       0.00257 &       222         \\ 
        K2      &       2014-11-19      &       201     &       60      &       224     &       2456981.35038   &       0.00332 &       287         \\ 
        K2      &       2014-11-20      &       203     &       60      &       246     &       2456982.12025   &       0.00384 &       332         \\ 
        K2      &       2014-11-21      &       325     &       60      &       252     &       2456982.92710   &       0.00318 &       275         \\ 
        K2      &       2014-11-22      &       199     &       60      &       252     &       2456983.71289   &       0.00270 &       233         \\ 
        K2      &       2014-11-23      &       200     &       60      &       253     &       2456984.50296   &       0.00505 &       437         \\ 
        K2      &       2014-11-23      &       205     &       60      &       250     &       2456985.29435   &       0.00330 &       285         \\ 
        K2      &       2014-11-24      &       200     &       60      &       228     &       2456986.08074   &       0.00329 &       285         \\ 
        K2      &       2014-11-25      &       166     &       60      &       266     &       2456986.85811   &       0.00602 &       520         \\ 
        K2      &       2014-11-26      &       203     &       60      &       273     &       2456987.65827   &       0.00467 &       404         \\ 
        K2      &       2014-11-26      &       490     &       60      &       254     &       2456988.45285   &       0.00316 &       273         \\ 
        K2      &       2014-11-27      &       199     &       60      &       255     &       2456989.24063   &       0.00251 &       216         \\ 
        K2      &       2014-11-28      &       200     &       60      &       279     &       2456990.02070   &       0.00245 &       211         \\ 
        K2      &       2014-11-29      &       147     &       60      &       240     &       2456990.82061   &       0.00474 &       410         \\ 
        K2      &       2014-11-30      &       204     &       60      &       234     &       2456991.60066   &       0.00448 &       387         \\ 
        K2      &       2014-11-30      &       200     &       60      &       260     &       2456992.40463   &       0.00489 &       423         \\ 
        K2      &       2014-12-01      &       202     &       60      &       274     &       2456993.18662   &       0.00362 &       313         \\ 
        K2      &       2014-12-02      &       201     &       60      &       261     &       2456993.97574   &       0.00593 &       512         \\ 
        K2      &       2014-12-03      &       177     &       60      &       263     &       2456994.75999   &       0.00508 &       439         \\ 
        K2      &       2014-12-04      &       408     &       60      &       239     &       2456995.56292   &       0.00331 &       286         \\ 
        K2      &       2014-12-04      &       205     &       60      &       241     &       2456996.34540   &       0.00296 &       256         \\ 
        K2      &       2014-12-05      &       202     &       60      &       262     &       2456997.14358   &       0.00266 &       230         \\ 
        K2      &       2014-12-06      &       200     &       60      &       259     &       2456997.92454   &       0.00275 &       237         \\ 
        K2      &       2014-12-07      &       201     &       60      &       247     &       2456998.71861   &       0.00327 &       283         \\ 
        K2      &       2014-12-08      &       436     &       60      &       250     &       2456999.51625   &       0.00350 &       302         \\ 
        K2      &       2014-12-08      &       198     &       60      &       246     &       2457000.29460   &       0.00373 &       322         \\ 
        K2      &       2014-12-09      &       202     &       60      &       254     &       2457001.08737   &       0.00246 &       213         \\ 
        K2      &       2014-12-10      &       201     &       60      &       236     &       2457001.87632   &       0.00405 &       350         \\ 
        K2      &       2014-12-11      &       204     &       60      &       281     &       2457002.66183   &       0.00265 &       229         \\ 
        K2      &       2014-12-11      &       202     &       60      &       244     &       2457003.45503   &       0.00373 &       323         \\ 
        K2      &       2014-12-12      &       562     &       60      &       271     &       2457004.25990   &       0.00611 &       528         \\ 
        K2      &       2014-12-13      &       199     &       60      &       283     &       2457005.03130   &       0.00306 &       264         \\ 
        K2      &       2014-12-14      &       198     &       60      &       249     &       2457005.81762   &       0.00296 &       255         \\ 
        K2      &       2014-12-15      &       201     &       60      &       234     &       2457006.60588   &       0.00432 &       373         \\ 
        K2      &       2014-12-15      &       201     &       60      &       253     &       2457007.39847   &       0.00258 &       223         \\ 
        K2      &       2014-12-16      &       200     &       60      &       259     &       2457008.19565   &       0.00477 &       412         \\ 
        K2      &       2014-12-17      &       200     &       60      &       254     &       2457008.98670   &       0.00362 &       313         \\ 
        K2      &       2014-12-18      &       201     &       60      &       230     &       2457009.76817   &       0.00270 &       233         \\ 
        K2      &       2014-12-19      &       198     &       60      &       267     &       2457010.55274   &       0.00578 &       499         \\ 
        K2      &       2014-12-19      &       201     &       60      &       257     &       2457011.34426   &       0.00246 &       212         \\ 
        K2      &       2014-12-20      &       330     &       60      &       252     &       2457012.13715   &       0.00429 &       370         \\ 
        K2      &       2014-12-21      &       204     &       60      &       255     &       2457012.93793   &       0.00681 &       589         \\ \hline
    \end{tabular}
    \tablefoot{The columns give the instrument name, the UT date closest to the transit center $T_0$, the number of photometric data points, the average cadence of the light curve in seconds, the scatter of the residuals from our best-fit model in parts per million, the transit time $T_0$ in the $\textrm{BJD}_\textrm{TDB}$ standard and its associated error, and the latter converted into seconds. The fitting procedure is described in detail in Section~\ref{sec:analysis_lc}.}
    \label{tab:tzero_e1}
\end{table*}

\begin{table*}
    \centering\renewcommand{\arraystretch}{1.3}
    \caption{Transits of planet WASP-47e included in our global analysis, part two.}
    \begin{tabular}{cccccccc} \hline\hline
        source & UT date & N & cadence & $\sigma_\mathrm{phot}$ & $T_0$ & $\sigma(T_0)$ & $\sigma(T_0)$ \\
        (+ band) & (of $T_0$) & & (s)  & (ppm) & ($\mathrm{BJD_{TDB}}$) & (days) & (s) \\ \hline
        K2      &       2014-12-22      &       202     &       60      &       251     &       2457013.71727   &       0.00428 &       370         \\ 
        K2      &       2014-12-23      &       203     &       60      &       245     &       2457014.50350   &       0.00308 &       266         \\ 
        K2      &       2014-12-23      &       437     &       60      &       278     &       2457015.29893   &       0.00384 &       331         \\ 
        K2      &       2014-12-24      &       483     &       60      &       253     &       2457016.08502   &       0.00295 &       255         \\ 
        K2      &       2014-12-25      &       201     &       60      &       265     &       2457016.87618   &       0.00274 &       237         \\ 
        K2      &       2014-12-26      &       200     &       60      &       255     &       2457017.65928   &       0.00437 &       377         \\ 
        K2      &       2014-12-26      &       200     &       60      &       263     &       2457018.43617   &       0.00806 &       696         \\ 
        K2      &       2014-12-27      &       199     &       60      &       252     &       2457019.24832   &       0.00340 &       294         \\ 
        K2      &       2014-12-28      &       202     &       60      &       263     &       2457020.03209   &       0.00206 &       178         \\ 
        K2      &       2014-12-29      &       202     &       60      &       254     &       2457020.81683   &       0.00758 &       654         \\ 
        K2      &       2014-12-30      &       202     &       60      &       264     &       2457021.61423   &       0.00370 &       320         \\ 
        K2      &       2014-12-30      &       203     &       60      &       272     &       2457022.40380   &       0.00212 &       183         \\ 
        K2      &       2014-12-31      &       201     &       60      &       238     &       2457023.19910   &       0.00341 &       294         \\ 
        K2      &       2015-01-01      &       204     &       60      &       234     &       2457023.98302   &       0.00273 &       236         \\ 
        K2      &       2015-01-02      &       203     &       60      &       260     &       2457024.76911   &       0.00598 &       517         \\ 
        K2      &       2015-01-03      &       197     &       60      &       277     &       2457025.55298   &       0.00472 &       407         \\ 
        K2      &       2015-01-03      &       201     &       60      &       254     &       2457026.35331   &       0.00363 &       314         \\ 
        K2      &       2015-01-04      &       201     &       60      &       258     &       2457027.13956   &       0.00633 &       547         \\ 
        K2      &       2015-01-05      &       202     &       60      &       278     &       2457027.92824   &       0.00362 &       313         \\ 
        K2      &       2015-01-06      &       574     &       60      &       267     &       2457028.81569   &       0.00465 &       402         \\ 
        K2      &       2015-01-07      &       426     &       60      &       276     &       2457029.53520   &       0.00262 &       226         \\ 
        K2      &       2015-01-07      &       199     &       60      &       281     &       2457030.31036   &       0.00386 &       334         \\ 
        K2      &       2015-01-08      &       200     &       60      &       256     &       2457031.09123   &       0.00237 &       205         \\ 
        K2      &       2015-01-09      &       202     &       60      &       241     &       2457031.87981   &       0.00380 &       329         \\ 
        K2      &       2015-01-10      &       705     &       60      &       277     &       2457032.69619   &       0.00495 &       427         \\ 
        K2      &       2015-01-10      &       360     &       60      &       268     &       2457033.46157   &       0.00282 &       243         \\ 
        K2      &       2015-01-11      &       199     &       60      &       254     &       2457034.24470   &       0.00566 &       489         \\ 
        K2      &       2015-01-12      &       194     &       60      &       244     &       2457035.03214   &       0.00453 &       391         \\ 
        K2      &       2015-01-13      &       180     &       60      &       255     &       2457035.82366   &       0.00243 &       210         \\ 
        K2      &       2015-01-14      &       204     &       60      &       244     &       2457036.61598   &       0.00198 &       171         \\ 
        K2      &       2015-01-14      &       200     &       60      &       262     &       2457037.39259   &       0.00715 &       618         \\ 
        K2      &       2015-01-15      &       200     &       60      &       260     &       2457038.19202   &       0.00411 &       355         \\ 
        K2      &       2015-01-16      &       202     &       60      &       213     &       2457038.98832   &       0.00280 &       242         \\ 
        K2      &       2015-01-17      &       170     &       60      &       238     &       2457039.77635   &       0.00498 &       430         \\ 
        K2      &       2015-01-18      &       201     &       60      &       246     &       2457040.56530   &       0.00370 &       320         \\ 
        K2      &       2015-01-18      &       471     &       60      &       253     &       2457041.35437   &       0.00278 &       240         \\ 
        K2      &       2015-01-19      &       202     &       60      &       261     &       2457042.14436   &       0.00263 &       227         \\ 
        K2      &       2015-01-20      &       199     &       60      &       270     &       2457042.93246   &       0.00312 &       270         \\ 
        K2      &       2015-01-21      &       169     &       60      &       258     &       2457043.72336   &       0.00304 &       263         \\ 
        K2      &       2015-01-22      &       199     &       60      &       251     &       2457044.50829   &       0.00253 &       218         \\ 
        K2      &       2015-01-22      &       365     &       60      &       257     &       2457045.30637   &       0.00452 &       390         \\ 
        K2      &       2015-01-23      &       203     &       60      &       261     &       2457046.08877   &       0.00422 &       365         \\ \hline
    \end{tabular}
    \tablefoot{The columns give the instrument name, the UT date closest to the transit center $T_0$, the number of photometric data points, the average cadence of the light curve in seconds, the scatter of the residuals from our best-fit model in parts per million, the transit time $T_0$ in the $\textrm{BJD}_\textrm{TDB}$ standard and its associated error, and the latter converted into seconds. The fitting procedure is described in detail in Section~\ref{sec:analysis_lc}.}
    \label{tab:tzero_e2}
\end{table*}

\end{appendix}

\end{document}